\documentclass[prd,aps,floats,floatfix,eqsecnum,nofootinbib]{revtex4}
\usepackage{verbatim,graphicx,amssymb,amsbsy,bm,amsmath,rotating,epsfig}
\usepackage{psfrag}
\usepackage{hyperref}
\usepackage{fontenc}
\newcommand{\be}{\begin{equation}}
\newcommand{\ee}{\end{equation}}
\newcommand{\bea}{\begin{eqnarray}}
\newcommand{\eea}{\end{eqnarray}}

\newcommand{\vx}{\ensuremath{\vec{x}}}
\newcommand{\va}{\ensuremath{{\vec \alpha}}}
\newcommand{\vk}{\ensuremath{\vec{k}}}
\newcommand{\vka}{\ensuremath{\vec{\kappa}}}
\newcommand{\vq}{\ensuremath{\vec{q}}}

\newcommand{\vQ}{\ensuremath{\vec{Q}}}

\begin{document}
\title{Cosmological evolution of warm dark matter fluctuations II: 
Solution from small to large scales and keV sterile neutrinos}
\author{\bf H. J. de Vega $^{(a,b)}$}
\email{devega@lpthe.jussieu.fr} 
\author{\bf N. G. Sanchez $^{(b)}$}
\email{Norma.Sanchez@obspm.fr} 
\affiliation{$^{(a)}$ LPTHE, Universit\'e
Pierre et Marie Curie (Paris VI) et Denis Diderot (Paris VII),
Laboratoire Associ\'e au CNRS UMR 7589, Tour 13-14, 4\`eme. et 5\`eme. \'etages, 
Boite 126, 4, Place Jussieu, 75252 Paris, Cedex 05, France. \\
$^{(b)}$ Observatoire de Paris,
LERMA, Laboratoire Associ\'e au CNRS UMR 8112.
 \\61, Avenue de l'Observatoire, 75014 Paris, France.}
\date{\today}
\begin{abstract}
We solve the cosmological evolution of warm dark matter (WDM) density fluctuations 
within the analytic framework of Volterra integral equations presented in the 
accompanying paper \cite{uno}. In the absence of neutrinos,
the anisotropic stress vanishes and the Volterra-type equations reduce 
to a single integral equation. We solve numerically this single 
Volterra-type equation both for DM fermions decoupling at thermal equilibrium 
and DM sterile neutrinos decoupling out of thermal equilibrium.
We give the exact analytic solution for the density fluctuations and gravitational 
potential at zero wavenumber. We compute the density contrast as a function of the scale factor $ a $
for a relevant range of wavenumbers $ k $. At fixed $ a $, the density 
contrast turns to grow with $ k $ for $ k < k_c $ while it decreases for $ k > k_c $, where 
$ k_c \simeq 1.6/$Mpc. The density contrast depends on $ k $ and $ a $ mainly through 
the product $ k \; a $ exhibiting a self-similar behavior. 
Our numerical density contrast for small $ k $ gently approaches our analytic solution for 
$ k = 0 $. For fixed $ k < 1/(60 \; {\rm kpc}) $, the density contrast generically 
grows with $ a $ while for $ k > 1/(60 \; {\rm kpc}) $ it exhibits 
oscillations starting in the radiation dominated (RD) era which become 
stronger as $ k $ grows. We compute the transfer function of the 
density contrast for thermal fermions and for sterile neutrinos decoupling out of 
equilibrium in two cases: the Dodelson-Widrow (DW) model and 
a model with sterile neutrinos produced by a scalar particle decay.
The transfer function grows with $ k $ for small $ k $ and 
then decreases after reaching a maximum at $ k = k_c $ reflecting the time evolution
of the density contrast. 
The integral kernels in the Volterra equations are nonlocal in time and 
their falloff determine the memory of the past evolution since decoupling. 
We find that this falloff is faster when DM decouples at thermal 
equilibrium than when it decouples out of thermal equilibrium. Although 
neutrinos and photons can be neglected in the matter dominated (MD) era,
they contribute to the Volterra integral equation in the MD era 
through their memory from the RD era.
\end{abstract}
\pacs{}
\keywords{DM, cosmological fluctuations}
\maketitle
\tableofcontents

\section{Introduction and Summary of Results}

In an accompanying paper \cite{uno} we provided a framework to study the 
complete cosmological evolution of dark matter (DM) density fluctuations for 
DM particles that decoupled being ultrarelativistic during the radiation 
dominated era which is the case of keV scale warm DM (WDM). 
In this paper, we solve the evolution of DM density fluctuations following the
framework developed in ref. \cite{uno}.

\medskip

The new framework presented in ref. \cite{uno} and here is generic for 
any type of DM and applies in particular to cold DM (CDM) too. 
The collisionless and linearized Boltzmann-Vlasov equations (B-V) 
for WDM and neutrinos in the presence of photons and coupled to the 
linearized Einstein equations are studied in detail in the presence of anisotropic 
stress with the Newtonian potential generically different from the spatial curvature 
perturbations. 

\medskip

In  ref. \cite{uno} the full system of B-V equations for DM and neutrinos 
is recasted as a system of coupled Volterra integral equations.
(Ref. \cite{bwu} has recently considered this issue in absence of anisotropic 
stress). These Volterra-type equations are valid both in the 
radiation dominated (RD) and matter dominated (MD) eras during which
the WDM particles are ultrarelativistic and then nonrelativistic.
This generalizes the so-called Gilbert integral equation only valid for 
nonrelativistic particles in the MD era. 

\medskip

We succeed to reduce the system of 
four Volterra integral equations for the density and anisotropic stress 
fluctuations of DM and neutrinos into a system of only two coupled Volterra 
equations. 

\medskip

In summary, the pair of partial differential Boltzmann-Vlasov
equations in seven variables for DM and for neutrinos become a system
of four Volterra linear integral equations on the density fluctuations 
$ \Delta_{dm}(\eta,\vk), \; \Delta_{\nu}(\eta,\vk) $ 
and anisotropic stress $ \Sigma_{dm}(\eta,\vk) \; , \Sigma_{\nu}(\eta,\vk) $
for DM and neutrinos, respectively. 

\medskip

In addition, because we deal with
linear fluctuations evolving on an homogeneous and isotropic cosmology, the 
Volterra kernel turns to be isotropic, independent of the  $ \vk $ directions.
As stated above, the $ \check{k} $ dependence factorizes out and we arrive to a final 
system of {\bf two} Volterra integral equations in two variables:  
the modulus $ k $ and the time that we choose to be as
\be\label{ay}
y \equiv  a(\eta)/a_{eq} \simeq 3200 \; a(\eta) \; .
\ee
We have thus considerably simplified the original problem: we reduce
a pair of partial differential B-V equations on seven 
variables $ \eta, \; \vq, \; \vx $
into a pair of Volterra integral equations on two variables: $ \eta, \; k $.

\medskip

The customary DM density contrast $ \delta(\eta,\vk) $ is 
connected with the density fluctuations $ \Delta_{dm}(\eta,\vk) $ by
\citep{mab}
\be\label{Ddi}
\delta(\eta,\vk) = \frac{\Delta_{dm}(\eta,\vk)}{\rho_{dm} \; 
[a_{eq} + a(\eta)]} \quad  ,  \quad a_{eq} \simeq \frac1{3200} \; ,
\ee
where $ \rho_{dm} $ is the average DM density today.

It is convenient to define dimensionless variables as 
$$
\alpha \equiv \frac{k \; l_{fs}}{\sqrt{I_4^{dm}}} \quad , \quad 
 l_{fs} = \frac2{H_0} \; \frac{T_d}{m} \; \sqrt{\frac{I^{dm}_4}{a_{eq} \; \Omega_{dm}}} \; ,
$$
where $ l_{fs} $ stands for the free-streaming length \cite{kt,bdvs,dvs},
$ T_d $ is the comoving DM decoupling temperature and $ I_4^{dm} $ is the 
dimensionless square velocity dispersion given by
\be\label{dfIn}
I_n^{dm}=\int_0^\infty Q^n \; f_0^{dm}(Q) \; dQ \quad ,  \quad {\rm while} \;  f_0^{dm}(Q) \; 
{\rm is ~normalized} \; 
 {\rm by} \quad I_2^{dm} = 1  \; .
\ee
$ Q $ is the dimensionless momentum $ Q \equiv q/T_d $ whose typical values are of order one.

\medskip

A relevant dimensionless rate emerges: the ratio between the DM particle mass $ m $ 
and the decoupling temperature at equilibration,
$$
\xi_{dm} \equiv \frac{m \; a_{eq}}{T_d} = 4900 \; \frac{m}{\rm keV} \; 
\left(\frac{g_d}{100}\right)^\frac13 \; ,
$$
$ g_d $ being the effective number of UR degrees of freedom at the DM decoupling.
Therefore, $ \xi_{dm} $ is a large number provided the DM is non-relativistic at equilibration.
For $ m $ in the keV scale we have $ \xi_{dm} \sim 5000 $. 

\medskip

DM particles and the lightest neutrino become non-relativistic by a redshift 
\be 
z_{trans} + 1 \equiv \frac{m}{T_d} \simeq 1.57 \times 10^7 \;   
\frac{m}{\rm keV} \; \left(\frac{g_d}{100}\right)^\frac13 \quad 
{\rm for ~ DM ~ particles}\quad ,
\quad z^{\nu}_{trans} = 34 \; \frac{m_{\nu}}{0.05 \; {\rm eV}}  \quad
{\rm for ~ the ~ lightest ~ neutrino} \; .
\ee
$ z_{trans} $ denoting the transition redshift from ultrarelativistic regime to the 
nonrelativistic regime of the DM particles.

\medskip

The final pair of dimensionless Volterra integral equations take the form
\bea\label{Ifinal}
&&{\breve \Delta}(y,\alpha) =  C(y,\alpha) + B_\xi(y)  \; {\bar \phi}(y,\alpha) +
\int_0^y dy' \left[G_\alpha(y,y') \; {\bar \phi}(y',\alpha) +
G^\sigma_\alpha(y,y') \; {\bar \sigma}(y',\alpha)\right] \; , \\ \cr
&& {\bar \sigma}(y,\alpha) =  C^\sigma(y,\alpha) + \int_0^y dy'  \left[
I^\sigma_\alpha(y,y') \;  {\bar \sigma}(y',\alpha) + I_\alpha(y,y') \; {\bar \phi}(y',\alpha)
\right] \; ,\label{Ifinal2}
\eea
with initial conditions $ {\breve \Delta}(0,\alpha) = 1 \quad , \quad {\bar \sigma}(0,\alpha) 
= \frac25 \; I_\xi \quad .$ 
This pair of Volterra equations is coupled with the linearized Einstein equations.

\medskip

The kernels and the inhomogeneous terms in eqs.(\ref{Ifinal})-(\ref{Ifinal2})
are given explicitly by eqs.(\ref{varnor})-(\ref{bynsi}), (\ref{asd4})-(\ref{uaysi})
and (\ref{gorda2})-(\ref{defialfa}). The arguments of these functions contain the 
dimensionless free-streaming distance $ l(y,Q) $,
\be\label{lfsI}
l(y,Q) = \int_0^y \frac{dy'}{\sqrt{\left[1+ y'\right] \; 
\left[ y'^2 + \displaystyle \left(\displaystyle Q/\xi_{dm}\right)^2 \right]}} \; .
\ee
\vskip -0.2 cm
The coupled Volterra integral equations (\ref{Ifinal})-(\ref{Ifinal2}) are easily amenable to a 
numerical treatment. 

\medskip

During the RD era the gravitational potential is dominated by the radiation 
fluctuations (photons and neutrinos).
The photons can be described in the hydrodynamical approximation 
(their anisotropic stress is negligible). 
The tight coupling of the photons to the electron/protons in the plasma 
suppresses before recombination all photon multipoles except
$ \Theta_0 $ and $ \Theta_1 $. (The  $ \Theta_l $ stem from the Legendre
polynomial expansion of the photon temperature fluctuations
$ \Theta(\eta, \vq, \vk) $ \cite{dod}). 

$ \Theta_0 $ and $ \Theta_1 $ obey the hydrodynamical equations \cite{dod}
\bea\label{hidro2}
&& \frac{d\Theta_0}{d \eta}  + k \; \Theta_1(\eta,\va) = \frac{d\phi}{d \eta} 
\quad , \\ \cr 
&& \frac{d\Theta_1}{d \eta} - \frac{k}3 \; \Theta_0(\eta,\va)= 
\frac{k}3 \;\phi(\eta,\va) \label{hidro} \quad . 
\eea
This is a good approximation for the purposes of following the DM 
evolution \cite{dod}.

\medskip

The photons gravitational potential
is given in the RD and MD eras by (ref. \cite{dod} and Appendix \ref{potr})
\be\label{firadI}
\phi(\eta,\vk) = \psi(\eta,\vk) = 3 \; \psi(0,\vk) \; 
\frac{\sqrt3}{\kappa \; y} \; j_1\left(\frac{\kappa \; y}{\sqrt3}\right) \quad , \quad
\kappa = k \; \eta^* \quad , \quad 
\eta^* \equiv \sqrt{\frac{a_{eq}}{\Omega_M}} \; \frac1{H_0} 
= 143 \; {\rm Mpc} \; ,
\ee
where $ j_1(x) $ is the spherical Bessel function of order one.

\medskip

\medskip

For redshift $ z < 30000 $ the kernel $ G_\alpha(y,y') $ in eq.(\ref{Ifinal}) simplifies as
\bea
&& G_\alpha(y,y')\buildrel{y, \, y' > 0.1}\over=  \frac{\xi_{dm} \; \kappa}{2 \, I_\xi} \; 
\frac{y \; y'}{\sqrt{1+y'}} \; \Pi\left[\alpha \;  \left( s(y) - s(y') \right)\right]  \quad 
{\rm where,}  \quad \cr \cr
&& \Pi(x) = \int_0^{\infty} Q \; dQ \; f_0^{dm}(Q) \; \sin( Q \; x)  \quad {\rm and} \quad 
s(y) = -{\rm Arg \, Sinh}\left( \displaystyle  \frac1{\sqrt{y}} \right) \; .\label{IPI}
\eea
In this regime  $ z < 30000 , \; y > 0.1 $, the anisotropic 
stress $ {\bar \sigma}(y,\alpha) $
turns to be negligible and eqs.(\ref{Ifinal})-(\ref{Ifinal2}) becomes a single Volterra integral equation.
In the MD era this equation takes the form
\be\label{Icmd}
\frac{{\breve \Delta}(y,\alpha)}{y} = g(y,\alpha) + \frac6{\alpha} 
\int_{s(1)}^{s(y)} ds' \; \Pi\left[\alpha\left(s(y)-s'\right)\right] \; {\breve \Delta}(y(s'),\alpha) 
\quad , \quad y \geq 1 \quad ,
\ee
where the inhomogeneous term $ g(y,\alpha) $ 
%is given by eq.(\ref{evf}) and 
contains the memory from the previous times $ y < 1 $ of the RD era. 
When DM is non-relativistic 
the memory from the regime where DM was ultrarelativistic turns out to fade out as 
$ 1/\xi_{dm} \sim 0.0002 $ compared to the recent memory where DM is non-relativistic. 

\medskip

The falloff of the kernel $ \Pi\left[\alpha\left(s-s'\right)\right] $ determines
the memory in the regime where DM is non-relativistic.
We find that this falloff is faster when DM decouples at thermal equilibrium
than when it decouples out of thermal equilibrium (see fig. \ref{pi}).
This can be explained by the general mechanism of thermalization \cite{ddv}:
in the out of equilibrium situation the momentum cascade towards the
ultraviolet is incomplete and there is larger occupation at 
low momenta and smaller occupation at large momenta than in the equilibrium 
distribution. Therefore, the out of equilibrium kernel $ \Pi(x) $ which is the Fourier
transform eq.(\ref{IPI}) of the freezed out momentum distribution
exhibits a longer tail than the equilibrium  kernel.

Neutrinos and photons can be neglected in the matter dominated era.
However, they contribute to the Volterra integral equation in the MD era 
through the memory integrals over $ 0 < y' < 1 $, namely the memory of the RD era.       

\medskip

When the anisotropic stress $ {\bar \sigma}(y, \alpha) $ is negligible,
eqs.(\ref{Ifinal})-(\ref{Ifinal2}) reduce to a single Volterra integral 
equation for the DM density fluctuations $ {\breve \Delta}_{dm}(y,\alpha) $ 
when the anisotropic stress  $ {\bar \sigma}(y, \alpha) $ is negligible.
We find the solution of this single Volterra equation for a broad range of 
wavenumbers $ 0.1 / {\rm Mpc} < k < 1/5 \; {\rm kpc} $.

\medskip

At zero wavenumber $ k = 0 $ the kernel of this Volterra equation vanishes 
and the DM fluctuations can be expressed explicitly in terms of the 
gravitational potential $ \phi $. The gravitational potential at $ k = 0 $ 
follows solely from the hydrodynamic equations for the radiation combined 
with the regularity requirement at $ k = 0 $ of the first linearized 
Einstein equation. Namely, the gravitational potential $ \phi $
is {\bf solely} obtained from the radiation without specifying the sources 
of the DM and radiation fluctuations. Using this explicit and well known 
form of $ \phi $,  (see e. g. ref. \cite{dod}) the DM fluctuations 
are obtained at $ \alpha = 0 $. The fact that the Einstein equations constrain 
their sources was first noticed in ref. \cite{eih} in a completely different
context.

\medskip

We depict in figs. \ref{flucd} the normalized density contrast vs. $ y $ 
(the scale factor divided by $ a_{eq} $) for thermal fermions and sterile 
neutrinos in the Dodelson-Widrow (DW) model \cite{dw} (both models yield
identical density fluctuations for a given value of $ \xi_{dm} $). 
Similar curves are obtained in the $ \chi $ model where sterile neutrinos 
are produced by the decay of a real scalar \cite{modelos}. 

\medskip

At fixed $ y $ we find that the density contrast grows with $ k $ for 
$ k < k_c $ while it decreases for $ k > k_c $, where 
%$ \alpha_c \simeq 0.1 $ which corresponds to 
$ k_c \simeq 1.6/$Mpc. We find that the density contrast 
depends on $ \alpha $ and $ y $ mainly through the product $ \alpha \; y $
exhibiting a self-similar behavior. 
The density contrast curves computed numerically for small $ \alpha $ gently
approach in the upper fig. \ref{flucd} our analytic solution for 
$ \alpha = 0 $. For fixed $ \alpha < 1 $, the density contrast generically 
grows with $ y $ while for $ \alpha > 1 $ it exhibits oscillations starting
in the RD era which become stronger as $ \alpha $ grows (see fig. 
\ref{flucd}). The density contrast becomes proportional to $ y $ (to the 
scale factor) at sufficiently late times. The larger is $ \alpha $, the 
later starts $ \delta(y,\alpha) $ to grow proportional to $ y $ (see fig. 
\ref{flucd}). Also, the larger is $ \alpha > 1 $, the later the 
oscillations remain.

\medskip

We depict in fig. \ref{talfa} the transfer function for thermal fermions
and sterile neutrinos in the DW model and for sterile neutrinos 
decoupling out of equilibrium in the $ \chi $ model.
The transfer function grows with $ k $ for small $ k $ and then decreases
after reaching a maximum at $ k = k_c $.

\medskip

We analyze in section \ref{vtunr} the system of two Volterra integral equations in the 
regimes where DM is in the transition from UR to NR and when DM is nonrelativistic. 
In sec. \ref{apef} we 
take the nonrelativistic limit of our system of Volterra integral equations in the MD era. 
This yields the Gilbert equation (plus extra terms). We find extra
memory terms and different inhomogeneities arising from our system of Volterra equations.

\medskip

In section \ref{numsol}
we consider the zero anisotropic stress case where the system of Volterra equations
reduces to a single Volterra equation. The numerical solution for the DM fluctuations
in a broad range of wavenumbers is presented and discussed, as well as the transfer function
and the analytic solution for zero wavenumber. 

\medskip

We present in sec. \ref{neuest} the distribution functions, main parameters and integral kernels
for sterile neutrinos decoupling out of equilibrium and compare
them to fermions decoupling with a Fermi-Dirac distribution.
Finally, we present in sec. \ref{apcdm} the generalization of the Volterra integral
equation for cold dark matter. 

\medskip

In the RD era where radiation fluctuations dominates the gravitational potential
we derive in Appendix \ref{potr} a second order differential equation for the gravitational potential.
We show that the solution of this differential equation is well approximated by the Bessel
function of order one eq.(\ref{firadI}). 

\medskip

We provide in Appendix \ref{apD} explicit and useful expressions for the free-streaming distance 
$ l(y,Q) $ [see eq.(\ref{lfsI})] in the main relevant regimes. 

\section{The Volterra Integral Equations and Relevant physical scales}

We recall here the pair of coupled Volterra integral equations derived in the accompanying
paper \cite{uno} from the Boltzmann-Vlasov equations for DM and for neutrinos.

\subsection{Density fluctuations and anisotropic stress fluctuations}

In the companion paper \cite{uno} we defined dimensionless density fluctuations 
$ {\bar \Delta}_{dm}(y,\alpha) $ and $ {\bar \Delta}_{\nu}(y,\alpha) $
and dimensionless anisotropic stress fluctuations $ {\bar \sigma}(y,\alpha) $
factoring out the initial gravitational potential $ \psi(0,\vk) $ 
in order to obtain quantities independent of the $ \vk $ direction.
These relevant quantities are expressed as 
\bea\label{defDb}
&& {\bar \Delta}_{dm}(y,\alpha)= \int \frac{d^3Q}{4 \, \pi} \; \varepsilon(y,Q) \;
f_0^{dm}(Q) \; \frac{\Psi_{dm}(y, \vQ, \vka)}{\psi(0,\vka)} \quad , \quad
{\bar \Delta}_{\nu}(y,\alpha)=\int \frac{d^3Q}{4 \, \pi} \; Q \;
f_0^{\nu}(Q) \; \frac{\Psi_{\nu}(y, \vQ, \vka)}{\psi(0,\vka)} \; , \cr \cr
&& \phi(y,\vka)=\psi(0,\vk) \; {\bar \phi}(y,\alpha) \quad , \quad 
\psi(y,\vka)=\psi(0,\vk) \; {\breve \psi}(y,\alpha) \quad {\rm and} \quad 
{\breve \psi}(0,\alpha) = 1 \; , \cr \cr
&& \sigma(y,\vka) = \psi(0,\vk) \; {\bar \sigma}(y,\alpha)
\quad , \quad {\bar \sigma}(y,\alpha) = {\bar \phi}(y,\alpha) - {\breve \psi}(y,\alpha)
\quad , \quad {\bar \sigma}(0,\alpha)={\bar \phi}(0,\alpha)-1 \; .
\eea
We then introduce in ref. \cite{uno} the combined density fluctuation $ {\breve \Delta}(y,\alpha) $
\be\label{dfDsom}
{\breve \Delta}(y,\alpha) = -\displaystyle\frac1{2 \, I_\xi} 
\left[ \displaystyle\frac1{\xi_{dm}} \; {\bar \Delta}_{dm}(y,\alpha) 
+  \displaystyle\frac{R_\nu(y)}{I_3^\nu} \; {\bar \Delta}_{\nu}(y,\alpha) \right]
\quad , \quad I_\xi = \displaystyle\frac{I_3^{dm}}{\xi_{dm}} + R_\nu(0) 
\simeq R_{\nu}(0) =  0.727 \quad , \quad {\breve \Delta}(0,\alpha) = 1 \; ,
\ee
where $ \xi_{dm} $ is the ratio between the DM particle mass $ m $ and the
physical decoupling temperature at equilibration redshift $ z_{eq} + 1 \simeq 3200 $,
\be\label{forxi}
\xi_{dm} = \frac{m \; a_{eq}}{T_d} = 4900 \; \frac{m}{\rm keV} \; 
\left(\frac{g_d}{100}\right)^\frac13 = 5520 \; \left(\frac{m}{\rm keV}\right)^\frac43 \; 
(g_{dm} \; N_{dm})^\frac13 \; .
\ee
We use here the dimensionless wavenumbers \cite{uno,bdvs}
\be\label{defxi}
\kappa \equiv k \; \eta^* \quad {\rm and} \quad
\alpha \equiv \frac2{\xi_{dm}} \; \kappa = 
\frac2{H_0} \; \frac{T_d}{m \; \sqrt{a_{eq} \; \Omega_{dm}}} \; k \; 
\quad {\rm where} \quad \eta^* \equiv \sqrt{\frac{a_{eq}}{\Omega_M}} \; 
\frac1{H_0} = 143 \; {\rm Mpc} \; .
\ee
Using $ {\breve \Delta}(y,\alpha) $ and $ {\bar \sigma}(y,\alpha) $ in 
ref. \cite{uno} allowed to reduce the system of four Volterra integral
equations into a the following pair of Volterra integral equations:
\bea\label{final}
&&{\breve \Delta}(y,\alpha) =  C(y,\alpha) + B_\xi(y)  \; {\bar \phi}(y,\alpha) +
\int_0^y dy' \left[G_\alpha(y,y') \; {\bar \phi}(y',\alpha) +
G^\sigma_\alpha(y,y') \; {\bar \sigma}(y',\alpha)\right] \; , \\ \cr\cr
&& {\bar \sigma}(y,\alpha) =  C^\sigma(y,\alpha) + \int_0^y dy'  \left[
I^\sigma_\alpha(y,y') \;  {\bar \sigma}(y',\alpha) + I_\alpha(y,y') \; 
{\bar \phi}(y',\alpha) \right] \; , \label{fsigma}
\eea
with the initial conditions \cite{uno}
$$ 
{\breve \Delta}(0,\alpha) = 1 \quad , \quad {\bar \sigma}(0,\alpha) = \frac25 \; I_\xi 
\simeq \frac25 \; R_\nu(0)\quad .
$$ 
%\vskip -0.5 cm
We have in eqs.(\ref{final})-(\ref{fsigma})
\bea\label{gorda2}
&& C(y,\alpha) = -\frac1{2 \, I_\xi} \; \left[ \frac{a(y,\alpha)}{\xi_{dm}} +\frac{R_\nu(y)}{I_3^\nu}\;
a^{ur}(y,\alpha)\right] \quad ,\quad  C^\sigma(y,\alpha)  \equiv \frac{a^\sigma(y,\alpha)}{\xi_{dm}} + 
\frac{R_\nu(y)}{I_3^\nu} \; a^{ur \; \sigma}(y,\alpha)\; , \label{calfa} \\ \cr %\cr
&& B_\xi(y) = -\frac1{2 \, I_\xi} \; \left[y \; b_{dm}(y) + 4 \, R_\nu(y)\right] \label{by}
\; , \cr \cr
&& G_\alpha(y,y') = -\frac{\kappa}{2 \, I_\xi \; \sqrt{1+y'} } \; 
\left[ \frac1{\xi_{dm}} \; N_\alpha(y,y') 
+ \frac{R_\nu(y)}{I_3^\nu} \; N^{ur}_\alpha(y,y') \right] 
\quad ,\label{galfa}  \\ \cr \cr   
&& G^\sigma_\alpha(y,y') =  -\frac{\kappa}{2 \, I_\xi \; \sqrt{1+y'}} \;
\left[\frac1{\xi_{dm}} \; N^\sigma_\alpha(y,y') 
- \frac{R_\nu(y)}{2 \, I_3^\nu} \; N^{ur}_\alpha(y,y') \right]  \label{gsalfa} \; , \\ \cr \cr 
&& I_\alpha(y,y') = \frac{\kappa}{\sqrt{1+y'}} \left[\frac1{\xi_{dm}} \; U_\alpha(y,y') + 
\frac{R_\nu(y)}{I_3^\nu} \; U^{ur}_\alpha(y,y')\right] \quad , \\ \cr \cr  
&& I^\sigma_\alpha(y,y')= \frac{\kappa}{\sqrt{1+y'}} \left[\frac1{\xi_{dm}} \; U^\sigma_\alpha(y,y') - 
\frac{R_\nu(y)}{2 \, I_3^\nu} \; U^{ur}_\alpha(y,y')\right] \; \label{defialfa}  \; .  
\eea
In eqs.(\ref{final})-(\ref{fsigma}) we can use $ I_\xi \simeq R_\nu(0) $. 
The DM integral kernels and inhomogeneity functions in eqs. 
(\ref{gorda2})-(\ref{defialfa}) are given by
\bea\label{varnor}
&& a(y,\alpha) =  \int_0^{\infty}  Q^2 \; dQ \; \varepsilon(y,Q)\left[
f_0^{dm}(Q) \; {\bar c}_{dm}^0(Q) + {\bar \phi}(0) \; \frac{df_0^{dm}}{d\ln Q}\right]
j_0\left[\frac{\alpha}2 \, Q \, l(y,Q) \right] \label{asd3} \; ,\\ \cr \cr
&& 
 y \; \xi_{dm} \; b_{dm}(y) =  \int_0^{\infty}\frac{Q^2 \; dQ}{\varepsilon(y,Q)}
\; f_0^{dm}(Q) \; \left[ 4 \, Q^2 + 3 \; (\xi_{dm} \; y)^2 \right] \label{dfb} \; ,\\ \cr \cr
&&  N_\alpha(y,y') =  \int_0^{\infty} Q^2 \; dQ \; \varepsilon(y,Q) \; 
\frac{df_0^{dm}}{dQ} \; j_1\left[\alpha \; l_Q(y,y') \right] \; 
\left[ \varepsilon(y',Q) +  \frac{Q^2}{\varepsilon(y',Q)}\right]  \label{byn}
\; ,\label{nucL2} \\ \cr \cr
&&  N_\alpha^{\sigma}(y,y') = -\int_0^{\infty} Q^2 \; dQ \; 
\frac{df_0^{dm}}{dQ} \; j_1\left[\alpha \; l_Q(y,y') \right] \;
\varepsilon(y,Q) \; \varepsilon(y',Q)   \; . \label{bynsi}\\ \cr \cr
&&  a^\sigma(y,\alpha) = \frac3{\kappa^2 \; y^2}\int_0^{\infty}  
\frac{Q^4 \; dQ}{\varepsilon(y,Q)} \left[
f_0^{dm}(Q) \; {\bar c}_{dm}^0(Q) + {\bar \phi}(0) \; \frac{df_0^{dm}}{d\ln Q}\right] \; 
j_2\left[\frac{\alpha}2 \, Q \, l(y,Q) \right] \label{asd4} \; , \\ \cr \cr
&&  U_\alpha(y,y')= -\frac3{5 \, \kappa^2 \; y^2} 
\int_0^{\infty}  \frac{Q^4 \; dQ}{\varepsilon(y,Q)} \; \frac{df_0^{dm}}{dQ}
\; \left[ \varepsilon(y',Q) +  \frac{Q^2}{\varepsilon(y',Q)}\right]
\left\{ 2 \; j_1\left[\alpha \; l_Q(y,y') \right]
- 3 \; j_3\left[\alpha \; l_Q(y,y')\right] \right\}\; ,  \label{uay} \\ \cr \cr
&&  U^{\sigma}_\alpha(y,y') = \frac3{5 \, \kappa^2 \; y^2} 
\int_0^{\infty}  \frac{Q^4 \; dQ}{\varepsilon(y,Q)} \; \frac{df_0^{dm}}{dQ} \; \varepsilon(y',Q)
\left\{ 2 \; j_1\left[\alpha \; l_Q(y,y')\right] - 3 \; j_3\left[\alpha \; l_Q(y,y')\right]\right\}
\; . \label{uaysi} 
\eea
The function $ {\bar c}_{dm}^0(Q) $ determines the intial conditions. We have for
thermal initial conditions (TIC) and for thermal initial conditions (TIC) \cite{uno}
\be\label{condini}
{\bar c}^0_{dm}(Q) =
\left\{\begin{array}{l} \displaystyle
\frac12 \;  \frac{d\ln f^{dm}_0}{d\ln Q} \quad 
{\rm for ~ thermal ~ initial ~ conditions ~ (TIC)} \; , \\ \\
-2 \quad {\rm for ~ Gilbert ~ initial ~ conditions ~ (GIC)} \quad .
\end{array} \right.
{\bar c}^0_\nu(Q) =
\left\{\begin{array}{l} \displaystyle
\frac12 \;  \frac{d\ln f^\nu_0}{d\ln Q} \quad 
{\rm for ~ TIC} \; , \\ \\
-2 \quad {\rm for ~ GIC} \quad .
\end{array} \right.
\ee
The neutrino integral kernels in eqs.(\ref{gorda2})-(\ref{defialfa}) and inhomogeneity functions 
are given by
\bea\label{rur}
&& N^{ur}_\alpha(y,y') = -8 \, I_3^{\nu} \; j_1\left[\kappa \; r(y,y')\right] \; , \cr \cr
&& U^{ur}_\alpha(y,y') = \frac{24 \, I_3^{\nu}}{5 \; \kappa^2 \; y^2} \; 
\left\{ 2 \; j_1\left[ \kappa \; r(y,y') \right] 
- 3 \;  j_3\left[ \kappa \; r(y,y') \right] \right\} \; , \\ \cr %\cr
&& a^{ur}(y,\alpha) = -2 \,  I_3^{\nu} \left[1 + 2 \; \bar \phi(0) \right] \;
j_0\left[ \kappa \; r(y,0)\right] \; \quad , \label{aur} \\ \cr 
&& a^{ur \; \sigma}(y,\alpha) = -6 \,  I_3^{\nu} \left[1 + 2 \; \bar \phi(0) \right] \;
\frac{j_2\left[\kappa \; r(y,0) \right]}{\kappa^2 \; y^2} 
\quad . \label{asur}
\eea
The Volterra integral equations (\ref{final})-(\ref{fsigma}) are coupled
with the linearized Einstein equations derived in the accompanying
paper \cite{uno}
\be\label{ecpgsd}
\left[\left(1+{\cal R}_0(y)\right)  \left( \frac{d}{dy} + 1  \right) +
\frac13 \left(\kappa \; y \right)^2 \right]{\bar \phi}(y,\alpha) 
= [1+{\cal R}_0(y)] \;  {\bar \sigma}(y,\alpha) -\frac1{2 \, \xi_{dm}} \;  
{\bar \Delta}_{dm}(y,\alpha) -\frac{R_\nu(y)}{2 \, I^\nu_3} \;  
{\bar \Delta}_\nu(y,\alpha) -2 \,  R_\gamma(y) \; {\bar \Theta}_0(y,\alpha) \; . 
\ee
Here,
\bea\label{defR}
&& {\cal R}_0(y) \equiv \frac{\rho_{dm}(y)}{\rho_r(y)} = 
\int_0^{\infty} Q^2 \; dQ \; \sqrt{y^2 + \frac{Q^2}{\xi_{dm}^2}} \;
f_0^{dm}(Q) \quad , \quad 
\rho_r(y) = \frac{\rho_r}{a^4(y)} \quad  {\rm and}\\ \cr\cr
&& {\cal R}_0(y) =\left\{\begin{array}{l} \displaystyle
\frac{I_3^{dm}}{\xi_{dm}}
\left[1 + {\cal O}\left(\xi_{dm}^2 \; y^2\right) \right]
\quad , \quad \xi_{dm} \; y \lesssim 1 \; , \\
y + \displaystyle \frac{I_4^{dm}}{2 \, \xi_{dm}^2 \; y} +  
{\cal O}\left(\frac1{\xi_{dm}^4 \; y^3}\right)\quad , 
\quad \xi_{dm} \; y \gtrsim 5 \; .
\end{array} \right.\label{R}
\eea

\subsection{Relevant scales in the ultra-relativistic and non-relativistic DM regimes}\label{escrele}

The evolution of the DM fluctuations presented here is valid generically 
for DM particles that decouple at redshift $ z_d $,
being ultrarelativistic in the RD era and
become non-relativistic in the same RD era. That is, the evolution 
presented here is valid as long as $ \xi_{dm} \gg 1 $
which is the case from  eq.(\ref{forxi}) provided DM decouples 
ultrarelativistically deep enough in the RD era. 

\medskip

The framework presented in this paper is general, valid for any DM particle,
not necessarily in the keV scale.
More precisely, the treatment presented here is valid for $ \xi_{dm} \gg 1 $ {\bf and}:
$$
1 \gg \frac{m}{T_{d \, phys}} = \frac{m}{T_d \; z_d} = 3200 \; \frac{\xi_{dm}}{z_d}
$$
which implies $ z_d \gg 3200 \; \xi_{dm} $.
The redshift at decoupling turns to be
\be
z_d + 1 = \frac{T_{d \, phys}}{T_d} = 1.57 \; 10^{15} \; 
\frac{T_{d \, phys}}{100 \; {\rm GeV}} \; 
\left(\frac{g_d}{100}\right)^\frac13 \; .
\ee
\vskip -0.2 cm
where we used $ T_d = (2/g_d)^{1/3} \; T_{cmb} $ and $ T_{cmb} = 0.2348 $ meV.

\medskip

DM particles are ultra-relativistic (UR) for
$ z \gtrsim z_{trans} , \;  z_{trans} $ being the redshift at the 
transition from ultra-relativistic to non-relativistic DM particles
\vskip -0.6 cm
\be 
z_{trans} + 1 \equiv \frac{m}{T_d} \simeq 1.57 \times 10^7 \;   
\frac{m}{\rm keV} \; \left(\frac{g_d}{100}\right)^\frac13 \; .
\ee
Then, they become non-relativistic (NR) for $ z \lesssim z_{trans} $. 
In terms of the variable
$ y $ [eq.(\ref{ay})] the transition from UR to NR DM particles takes place
around $ y \sim y_{trans} $ while decoupling happens well before $ y_{trans} $
by $ y \sim y_d $:
$$
y_{trans} = 1/\xi_{dm} \simeq 0.0002 \quad , \quad
y_d = 3200/z_d \simeq 2 \times 10^{-12} \; .
$$
Notice that modes that reenter the horizon by the UR-NR transition
$ y \sim y_{trans} $, have from eqs. (2.30) and (2.40) of the accompanying paper
ref. \cite{uno}, wavenumbers
$$
k \sim \frac1{\eta^* \; y_{trans}} \sim \frac{\xi_{dm}}{\eta^*}
= \frac{2 \; \sqrt{I_4^{dm}}}{l_{fs}}\sim  \frac1{l_{fs}} \; .
$$
That is, when DM particles become nonrelativistic the free-streaming length
$ l_{fs} $ is of the order of the comoving horizon \cite{bwu}.

\begin{table}
\begin{tabular}{|c|c|c|} \hline  
 & & \\
Universe Event & redshift $ z $ & $ y = \displaystyle \frac{a}{a_{eq}}=\displaystyle
\frac{z_{eq}+1}{z+1} \simeq \frac{3200}{z+1}$ \\ 
 & & \\
\hline \hline
 & & \\
DM decoupling & $z_d \sim 1.6 \; 10^{15} \; \frac{T_{dp}}{100 \; {\rm GeV}} \; 
\left(\frac{g_d}{100}\right)^\frac13$ & $ y_d \simeq 2 \times 10^{-12}$  \\ 
 & & \\ \hline 
 & & \\
 neutrino decoupling & $z^{\nu}_d \simeq 6 \times 10^9$ & $ y^{\nu}_d \simeq 0.5 \times 10^{-6} $ \\
 & & \\ \hline 
 & & \\
DM particles transition from UR to NR & $z_{trans} \simeq 1.6 \times 10^7 \; \frac{\rm keV}{m} \;
\left(\frac{g_d}{100}\right)^\frac13$ & $y_{trans} = \displaystyle \frac1{\xi_{dm}} \simeq 0.0002$ \\
 & &  $ 10^{-6} < y < 0.01 $ \\ \hline
 & & \\
Transition from the RD to the MD era & $ z_{eq} \simeq 3200 $ & $ y_{eq} = 1 $ \\
 & & \\ \hline   
 & & \\
The lightest neutrino becomes NR & $z^{\nu}_{trans} = 95 \; \displaystyle \frac{m_{\nu}}{0.05 \; {\rm eV}}$ & 
$y^{\nu}_{trans} = 34 \; \displaystyle \frac{0.05 \; {\rm eV}}{m_{\nu}}$ \\
 & & \\ \hline   
 & & \\
Today & $ z_0 = 0 $ & $ y_0 \simeq 3200 $  \\
 & & \\ \hline  
\end{tabular}
\caption{Main events in the DM, neutrinos and universe evolution.}
\label{esca}
\end{table}

\medskip

At decoupling, the covariant neutrino temperature, decoupling neutrino 
redshift and $ y $ variable are,
$$ 
T_d^{\nu}= 0.17 \; 10^{-3} \; {\rm eV} \quad ,  \quad  z^{\nu}_d \simeq 6 \times 10^9 \quad 
{\rm and}  \quad y^{\nu}_d \simeq 0.5 \times 10^{-6}  \; .
$$
The lightest neutrinos become non-relativistic at a redshift 
$$
z^{\nu}_{trans} = 95 \; \frac{m_{\nu}}{0.05 \; {\rm eV}} \quad {\rm and}  \quad 
y^{\nu}_{trans} = 34 \; \frac{0.05 \; {\rm eV}}{m_{\nu}} \; .
$$
Namely, neutrinos become non-relativistic in the MD era when their 
density as well as their fluctuations are negligible. Thus, we can
treat the neutrinos as ultra-relativistic or neglect them. 

\medskip

The neutrino and photon fractions of the energy density are defined in general as
$$
R_\nu(\eta) \equiv \frac{\rho_\nu(\eta)}{\rho(\eta)} = \frac{\Omega_\nu}{\Omega_r + a(\eta) \; \Omega_M}
\quad , \quad
R_\gamma(\eta) \equiv \frac{\rho_\gamma(\eta)}{\rho(\eta)} = 
\frac{\Omega_\gamma}{\Omega_r+ a(\eta) \; \Omega_M}
$$
where $ \rho_\nu(\eta), \; \rho_\gamma(\eta) $ and $ \rho(\eta) $ stand
for the neutrino, photon and total energy density, respectively. In the radiation
dominated era $ \Omega_r \gg a(\eta) \; \Omega_M $ and $ R_\nu(\eta) + R_\gamma(\eta) = 1 $.
The neutrino fraction changes after neutrino decoupling when the cosmic temperature 
crosses the $ e^+-e^- $ threshold, that is \cite{dod}, 
\be\label{rneta}
R_\nu(\eta)=\left\{\begin{array}{l} 0.727 \quad , \quad  4 \times 10^9 \lesssim z \lesssim 6 \times 10^9\\
0.41 \quad , \quad 3200 \lesssim z \lesssim 4 \times 10^9 \\
0 \quad , \quad 0 \leq z \lesssim 3200
\end{array} \right.   \quad .
\ee
The quantity $ I_\xi $ defined by eq.(\ref{dfDsom}) is dominated by 
the neutrino piece $ R_{\nu}(0) $ and takes the value
\be\label{itzi}
I_\xi \simeq R_{\nu}(0) =  0.727 \; . 
\ee
In the MD dominated era both $ R_\nu(\eta) $ and $ R_\gamma(\eta) $
become very small and can be neglected.

\medskip

We summarize in Table \ref{esca} the ranges of the redshift 
$ z $ and the variable $ y $ (the scale factor normalized to
unity at equilibration) for the main events in the DM, 
neutrinos and the universe evolution.

\medskip

The free-streaming distance $ l(y,Q) $ is expressed by eq. (\ref{lfsI}).
$ l(y,Q) $ can be expressed in general in terms of elliptic
integrals. In the present case where $ \xi_{dm} \sim 5000 $
we find in appendix B excellent approximations to $ l(y,Q) $ in terms
of simple elementary functions. We display the free-streaming distance $ l(y,Q) $
and the particle energy $ \varepsilon(y,Q) $
for the different regimes in Table \ref{roT}. It must be stressed that
each of the four formulas displayed in Table \ref{roT} {\bf match} with 
its neighboring expression as discussed in appendix B.

\begin{table}
\begin{tabular}{|c|c|c|} \hline  
Range of Validity & $\varepsilon(y,Q)$ & $ l(y,Q) $  \\ \hline \hline
UR DM particles  & &  \\
$ \xi_{dm} \; y \ll 1 $ & $ Q $ & $ \displaystyle \frac{\xi_{dm} \; y}{Q} \; 
\frac1{\sqrt{1 +\frac{I_3^{dm}}{\xi_{dm}}}} $ \\ 
 $ 0 < y < 10^{-6} $ & & \\ \hline 
 Transition regime from UR & & \\
to NR DM particles & $\sqrt{Q^2 + (\xi_{dm})^2 \; y^2 }$ & 
$  \displaystyle {\rm Arg \, Sinh}\left(\displaystyle\frac{\xi_{dm} \; y}{Q}\right) $  \\
$ 10^{-6} < y < 0.01 $ & &  \\ \hline 
 NR DM particles & &  \\
$ 0.01 < y < 3200 $ & $ \xi_{dm} \; y $ & $ \displaystyle - 2 \; {\rm Arg \, Sinh}\left( \frac1{\sqrt{y}} \right) +
\displaystyle \log\left(\displaystyle \frac{8 \; \xi_{dm}}{Q} \right) +
\frac12 \; \frac{Q}{\xi_{dm}} - \frac18 \; \left(\frac{Q}{y \; \xi_{dm}}\right)^2
\left[3 \, y \; \sqrt{1+y} + y + 2 \right]$   \\
$ \xi_{dm} \; y \gg 1 $ & & \\ \hline   
 MD era  & &  \\
$ y \gg 1 $ & $ \xi_{dm} \; y $ & $ - \displaystyle\frac2{\sqrt{y}} +
\displaystyle \log\left(\displaystyle\frac{8 \; \xi_{dm}}{Q}\right)
+ \frac12 \; \frac{Q}{\xi_{dm}} $  \\ 
 NR DM particles  & & \\ \hline  
\end{tabular}
\caption{The different regimes ultra-relativistic (UR), transition and non-relativistic (NR)
of the free-streaming distance $ l(y,Q) $.
Notice that the second (third) formula for $ l(y,Q) $ is also valid
in the first (fourth) formula for $ 0 < y < 10^{-6} $ ($ y \gg 1 $).
In addition, the third formula of $ l(y,Q) $ for $ y \ll 1 $ matches for
$ \xi_{dm} \; y \gg 1 $ with the asymptotic behaviour of the second formula for $ l(y,Q) $.
The precise behaviours of $ l(y,Q) $ are derived in Appendix \ref{apD}
and given by eqs.(\ref{aprox1})-(\ref{roapr}). When DM is UR $ l(y,Q) $
grows as the comoving horizon $ \eta^* \, y $ and thus free-streaming
efficiently erases fluctuations. When DM becomes NR $ l(y,Q) $ grows
much slower and free-streaming is inefficient to erase fluctuations.}
\label{roT}
\end{table}

\medskip

From eqs.(\ref{forxi}) and (\ref{defxi}) we obtain for the dimensionless variable $ \alpha $, 
\be
\alpha = 58.37 \; \frac{\rm keV}{m} \; 
\left(\frac{100}{g_d}\right)^\frac13 \; k \; {\rm kpc} \; .
\ee
In terms of $ \alpha $,
the primordial gravitational potential eq.(3.17) in the accompanying paper \cite{uno} becomes,
\be\label{potpri}
\psi(0,\va) = \frac{1.848}{\alpha^\frac32} \; 
\left(\frac{\alpha}{\alpha_0}\right)^{\frac12(n_s-1)} \; 
\left(\frac{\rm keV}{m}\right)^3 \; \frac{100}{g_d} \; ({\rm kpc})^3 \; G(\va) \; ,
\ee
where $ \alpha_0 = 1.167 \; 10^{-4} \; ({\rm keV}/m) \; \left(100/g_d\right)^\frac13 $ and 
$$
< G(\va) \; G^*(\va') > = \delta(\va-\va') \; .
$$

\section{From the ultrarelativistic to the non-relativistic regime
of the DM in the Volterra equations}\label{vtunr}

We investigate here the system of Volterra integral equations 
(\ref{final})-(\ref{fsigma}) first in the transition regime for DM and 
then in the non-relativistic DM regime.

\begin{figure}[h]
\begin{center}
\begin{turn}{-90}
\psfrag{"fsq01.dat"}{$ l(y,Q=0.1)/Q $ vs. $ \log_{10} \; y $}
\psfrag{"fsq1.dat"}{$ l(y,Q=1)/Q $ vs. $ \log_{10} \; y $}
\psfrag{"fsq10.dat"}{$ l(y,Q=10)/Q $ vs. $ \log_{10} \; y $}
\includegraphics[height=13.cm,width=8.cm]{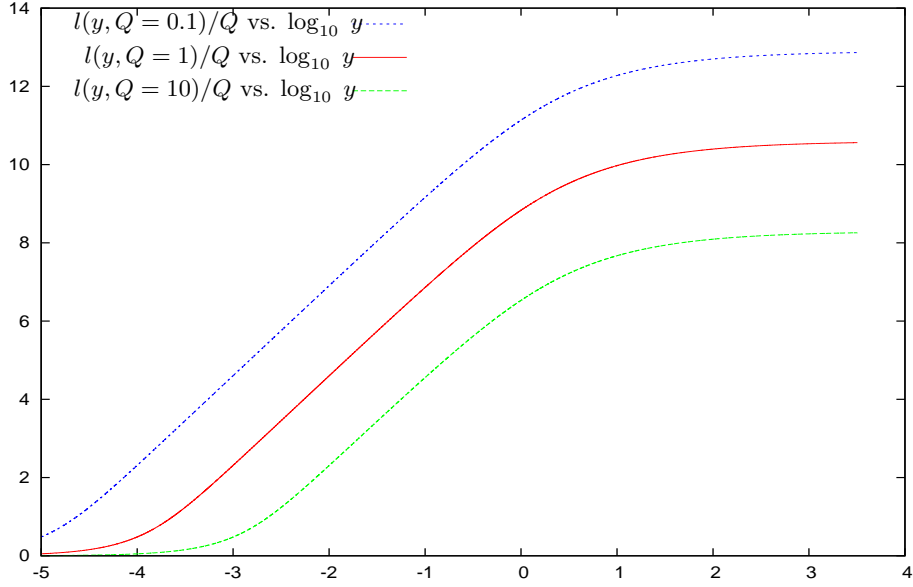}
\end{turn}
\caption{The free-streaming length in dimensionless variables
$ l(y,Q) $  divided by $ Q $ vs. $ \log_{10} \; y $ for $ Q = 0.1, \; 1 $ and
$ 10 $. [Recall that $ \lambda_{FS} = (\eta^*/\xi_{dm}) \; Q \;  l(y,Q) $ 
\cite{uno}]. We explicitly compute $ l(y,Q) $ in appendix \ref{apD}.
$ l(y,Q) $ is given in the different regimes by eqs.(\ref{lfsexa}), 
(\ref{aprox1}), (\ref{roapr}) and (\ref{eliplio}). Notice that 
$ \log_{10} y = 0 $ corresponds to equilibration. 
We choose here $ \xi_{dm} = 5000 $.} 
\label{roL}
\end{center}
\end{figure}

\subsection{Transition Regime}

We consider here the coupled Volterra integral equations 
(\ref{final}-(\ref{fsigma})) in the transition regime
from ultrarelativistic to non-relativistic DM particles $  0.5 \; 10^{-6} < y < 0.01 $
well inside the RD era where the neutrinos are ultrarelativistic and they have already decoupled.

\medskip

The second entry of Table \ref{roT} the one-particle energy 
$ \varepsilon(y,Q) =\sqrt{(\xi_{dm})^2 \; y^2 + Q^2}$ and the 
free-streaming length $ l(y,Q) $ applies now. Therefore, we have from eq.(\ref{aprox1}),
$$
l(y,Q) =  \left[1 -\frac3{16} \; \left(\frac{Q}{\xi_{dm}}\right)^2 \right]
\displaystyle {\rm Arg \, Sinh}\left(\displaystyle\frac{\xi_{dm} \; y}{Q}\right) 
- \frac12 \; \left[\left(1-\frac38 \; y \right)\sqrt{y^2 + \left(\frac{Q}{\xi_{dm}}\right)^2}
- \frac{Q}{\xi_{dm}}\right] + {\cal O}(y^3) \; . 
$$
$$
l_Q(y,y') = \frac{Q}2 \; \left[1 -\frac3{16} \; \left(\frac{Q}{\xi_{dm}}\right)^2 \right]
\; {\rm Arg \, Sinh}\left(\displaystyle\frac{\xi_{dm} \; y}{Q}\right)
%- {\rm Arg \, Sinh}\left(\displaystyle\frac{\xi_{dm} \; y'}{Q}\right)\right] 
- \frac{Q}4 \; \left(1-\frac38 \; y \right)\sqrt{y^2 + 
\left(\frac{Q}{\xi_{dm}}\right)^2} - \{ y \Rightarrow y'\} \quad .
$$
These formulas are to be inserted in eqs.(\ref{asd3})-(\ref{uaysi})
for $ a(y,\alpha) , \; a^\sigma(y,\alpha) , \;
%and in eqs.(\ref{nucL2})-(\ref{uaysi})
N_\alpha(y,y'), \; N^\sigma_\alpha(y,y'), \; U_\alpha(y,y') $ 
and $ U^{\sigma}_\alpha(y,y') $.

\subsection{Non-relativistic Regime}\label{reginr}

We write here the Volterra integral equation in the non-relativistic regime $ 3200 > y > 0.01 $.

\medskip

The third entry of Table \ref{roT} 
for the one-particle energy $ \varepsilon(y,Q) \simeq \xi_{dm} \; y $ and the free-streaming length
$ l(y,Q) $ applies in this case. Notice that the difference of the free-streaming lengths
which appears in the integrand of the kernel $ N_\alpha(y,y') $  eq.(\ref{byn})
is  now $Q$-independent because the DM particles are non-relativistic:
$$
l_Q(y,y') = \frac{Q}2 \; \left[l(y,Q)-l(y',Q) \right] = Q \; \left[s(y)-s(y') \right] \; ,
%{\rm Arg \, Sinh}\left( \frac1{\sqrt{y'}} \right) - {\rm Arg \, Sinh}\left( \frac1{\sqrt{y}} \right) 
$$
where we used eq.(\ref{roapr}), neglected terms 
$ {\cal O}\left( [Q/\xi_{dm}]^2 \; \log Q/\xi_{dm}\right) $ 
in the free-streaming length $ l(y,Q) $ and 
\be \label{defsy}
%\Sigma(y) \equiv \frac12 \; \log\frac{\sqrt{1+y} + 1}{\sqrt{1+y} - 1} =
s(y) \equiv -{\rm Arg \, Sinh}\left( \displaystyle  \frac1{\sqrt{y}} \right) \quad , \quad
\frac{ds}{dy} = \frac1{2 \; y \; \sqrt{1+y}} \quad .
%\quad s(y) \buildrel{y \gg 1}\over=-\frac1{\sqrt{y}} \; .
\ee
In the non-relativistic regime the kernels $ N_\alpha(y,y') $ and $ N_\alpha^\sigma(y,y') $ 
in eqs.(\ref{byn})-(\ref{bynsi}) both in the RD and the MD eras become,
$$
N_\alpha(y,y') =  -N_\alpha^\sigma(y,y') = (\xi_{dm})^2 \; y \; y' \int_0^{\infty} Q^2 \; dQ \;
\frac{d f_0^{dm}}{dQ} \;  
%j_1\left\{\alpha \; Q \left[ {\rm Arg \, Sinh}\left( \frac1{\sqrt{y'}} \right) 
%- {\rm Arg \, Sinh}\left( \frac1{\sqrt{y}} \right) \right]\right\} 
j_1\left\{\alpha \; Q \left[s(y)-s(y')\right]\right\} \; .
$$
Integrating by parts $ df_0^{dm}/dQ $ in the above integral leads to the simpler expression
\be\label{Nnr}
N_\alpha(y,y') = -N_\alpha^\sigma(y,y') = - (\xi_{dm})^2 \; y \; y' \; 
\Pi\left[ \alpha \; \left( s(y)-s(y')\right)\right] \quad {\rm for} \quad y, \; y' > 0.01 \; ,
\ee
where
\be\label{defpi}
\Pi(x) \equiv \int_0^{\infty} Q \; dQ \; f_0^{dm}(Q) \; \sin( Q \; x) \; .
\ee
That is, in the nonrelativistic regime the kernel $ N_\alpha(y,y') $ becomes the Fourier
transform of the zeroth order momentum distribution $ f_0^{dm}(Q) $. Notice that 
$$
\Pi(0) = 0 \quad , \quad \Pi'(0) = 1 \quad ,
$$
where we used eqs.(\ref{dfIn}) and (\ref{defpi}).

\medskip

In a similar way we obtain for the kernels $ U_\alpha(y,y') $ and $ U_\alpha^\sigma(y,y') $
[given by eqs.(\ref{uay})-(\ref{uaysi})] in the nonrelativistic regime, 
\be
 U_\alpha(y,y')= -U_\alpha^\sigma(y,y')=-\frac{3 \, y'}{5 \, \kappa^2 \; y^3} \int_0^{\infty} \; 
Q^4 \; dQ \; \frac{df_0^{dm}}{dQ} \left\{ 2 \; j_1\left(\alpha \; Q \; \left[s(y)-s(y')\right]
\right) - 3 \; j_3\left(\alpha \; Q \; \left[s(y)-s(y')\right]\right) \right\} \; .
\ee
Upon integrating by parts this formula can be recasted in the simpler form
$$
U_\alpha(y,y')= -U_\alpha^\sigma(y,y')=
\frac{3 \, \alpha \; \left[s(y)-s(y')\right] \; y'}{\kappa^2 \; y^3} \int_0^{\infty} \; 
Q^4 \; dQ \; f_0^{dm}(Q) \; \left\{ j_0\left(\alpha \; \left[s(y)-s(y')\right] \; Q\right)
- \frac{j_1\left(\alpha \; \left[s(y)-s(y')\right] \; Q\right)}{\alpha \; 
\left[s(y)-s(y')\right] \; Q} \right\} \; .
$$
where we used also the angular integrals in the Appendix B of the accompanying paper ref. \cite{uno}.

\medskip

When the DM particles become nonrelativistic the anisotropic stress 
$ {\bar \sigma}(y,\alpha) $ decreases fast as $ 1/ (\kappa \; y)^2 $. 
For $ y \; \alpha \gtrsim 1 $, then $ 1/ (\kappa \; y)^2 < 10^{-6} $ 
 the anisotropic stress can be neglected. Therefore, for 
$ y \; \alpha \gtrsim 1 $ the system of Volterra equations 
(\ref{final})-(\ref{fsigma}) 
reduces to a single Volterra equation for the density fluctuations
$ {\breve \Delta}(y,\alpha) $.

\medskip

In this nonrelativistic regime where $ \kappa \; y \gg 1 , \; \varepsilon(y,Q) \simeq \xi_{dm} \; y $, 
the inhomogeneous pieces $ C(y,\alpha) $ and $ C^\sigma(y,\alpha) $ from eqs.(\ref{asd3}), (\ref{asd4}), 
(\ref{aur}), (\ref{asur}) and (\ref{gorda2}) become
\bea\label{gnr}
&& C(y,\alpha) = -\frac{y}{2 \, I_\xi} \; \int_0^{\infty} Q^2 \; dQ \; 
\left[ f_0^{dm}(Q) \; {\bar c}_{dm}^0(Q)+{\bar \phi}(0) \; \frac{df_0^{dm}}{d\ln Q}\right]
j_0\left[\frac{\alpha}2 \, Q \; l^{NR}(y,Q) \right] +  \cr \cr\cr
&& +  \frac{R_\nu(y)}{I_\xi} \; \left[1 + 2 \, {\bar \phi}(0)\right]
\; j_0\left[\kappa \; r(y,0) \right] \; ,\cr\cr\cr
&& C^{\sigma}(y,\alpha) = \frac3{(\kappa \; \xi_{dm})^2 \; y^3} \int_0^{\infty}  
Q^4 \; dQ \;\left[ f_0^{dm}(Q) \; {\bar c}_{dm}^0(Q)+{\bar \phi}(0) \; \frac{df_0^{dm}}{d\ln Q}\right]
\; j_2\left[\frac{\alpha}2 \, Q \; l^{NR}(y,Q) \right]  \cr\cr\cr
&& - 6 \; R_\nu(y) \; \left[1 + 2 \, {\bar \phi}(0)\right]
\; \frac{j_2\left[\kappa \; r(y,0) \right]}{\kappa^2 \; y^2} \; ,
\eea
where we used eq.(\ref{roapr})
\be\label{lnr}
l(y,Q) \simeq l^{NR}(y,Q) \equiv 
2 \; s(y) + \log\left(8 \; \xi_{dm}/Q\right) \; .
\ee  
The Volterra equation (\ref{final})-(\ref{fsigma}) at $ y $ involves the 
integral over all  $ y' $ in the interval $ 0 < y' < y $. Namely, we need the kernel
$ N_\alpha(y,y') $ for all  $ y' $ in $ 0 < y' < y $. Therefore, in the 
nonrelativistic regime $ y > y_1 = 0.01 $ we need the kernels with mixed arguments,
where $ y'$ belongs to the transition or to the ultrarelativistic regime 
($ 0 < y' < y_1 $).

\medskip

We obtain from eq.(\ref{byn}) for the mixed kernel
\bea \label{mixto}
&&N_\alpha(y,y') =  \xi_{dm} \; y \int_0^{\infty} Q^2 \; dQ \; \frac{d f_0^{dm}}{dQ} \;
 j_1\left\{\frac{\alpha}2 \; Q \; \left[\log\left(\frac{8 \; \xi_{dm}}{Q}\right) - 2 \; 
{\rm Arg \, Sinh}\left( \frac1{\sqrt{y}} \right)-
{\rm Arg \, Sinh}\left(\displaystyle\frac{\xi_{dm} \; y'}{Q}\right)\right]\right\} \times \cr \cr
&& \times \left[ \varepsilon(y',Q) +  \frac{Q^2}{\varepsilon(y',Q)}\right] 
\quad  {\rm for} \quad y > y_1 =0.01 \quad , \quad y' < y_1 =0.01 \; . 
\eea

The kernel $ N_\alpha(y,y') $ is proportional to $ (\xi_{dm})^2 $ when both $ y $ and $ y' $
are in the nonrelativistic regime [eq.(\ref{Nnr})] while it is proportional to $ \xi_{dm} $ when
$ y $ is  in the nonrelativistic regime and $ y' $ is in the 
transition or ultrarelativistic regimes [eq.(\ref{mixto})]. 
Namely, in the nonrelativistic regime,
the memory of the transition regime and ultrarelativistic regime fades as $ 1/\xi_{dm} \sim 0.0002 $.

\medskip

In the MD dominated era $ y > 1 $, radiation (photons and neutrinos) can be 
neglected: $ R_\nu(y)= R_\gamma(y)=0 $. Once neutrinos are negligible, 
the anisotropic stress $ {\bar \sigma}(y,\kappa) $
becomes very small and can be neglected too. Therefore, we have for $ y > 1 $ 
dropping the neutrino contributions in eqs.(\ref{dfDsom}), (\ref{final}) and 
(\ref{gorda2})-(\ref{galfa}),
\bea\label{volmd}
&& {\bar \Delta}_{dm}(y,\kappa) = -2 \, \xi_{dm} \; I_\xi \; {\breve \Delta}(y,\kappa) \quad , \quad \qquad \qquad
{\rm MD ~ era} \\ \cr
&& {\bar \Delta}_{dm}(y,\alpha)=   a(y,\alpha) + y \; \xi_{dm} \; 
b_{dm}(y) \; {\bar \phi}(y,\alpha)  + \kappa \; \int_0^y \frac{dy'}{\sqrt{1+y'}} \;
N_\alpha(y,y')\; {\bar \phi}(y',\alpha)+\kappa \; \int_0^1 \frac{dy'}{\sqrt{1+y'}} \;
N_\alpha^{\sigma}(y,y') \; {\bar \sigma}(y',\alpha)  \; \; . \nonumber
\eea
Notice that the integrals here cover the radiation dominated era $ 0 < y < 1 $.
That is, the memory of the neutrinos and photons during the RD era is preserved in the MD era.

The linearized Einstein equation for the gravitational potential in the MD era
$ {\bar \phi}(y,\alpha) = {\bar \psi}(y,\alpha) $ become from eq.(\ref{ecpgsd})
\be\label{eilmd}
\left[y(1+y) \; \frac{d}{dy} + 1 + y +\frac13 \left(\kappa \; y \right)^2 \right]{\bar \phi}(y,\alpha) 
= -\frac1{2 \, \xi_{dm}} \;  {\bar \Delta}_{dm}(y,\alpha) \quad .
\ee
This equation can be solved as
\be\label{fidm}
{\bar \phi}(y,\alpha) =-\frac1{2 \, \xi_{dm} \; y} \; 
\int_0^y \frac{dy'}{1+y'} \; \beta_\kappa (y,y') \;  {\bar \Delta}_{dm}(y',\alpha) 
\quad {\rm where} \quad 
\beta_{\kappa} (y,y') = \left(\frac{1+y}{1+y'} \; e^{y'-y} \right)^{\kappa^2 /3} \; .
\ee
We compute in appendix A of the accompanying paper \cite{uno} this integral 
in the asymptotic regime $ \kappa \; y \gg 1 $. We find at leading order 
from eq.(A2) of ref.\cite{uno},
\be\label{potmd}
{\bar \phi}(y,\alpha) \buildrel{\kappa \; y \gg 1}\over=-\frac3{2 \, \xi_{dm} \; (\kappa \; y)^2}
\; {\bar \Delta}_{dm}(y,\alpha) \; ,
\ee
which corresponds to the Poisson's law. This result applies for 
$ \kappa = \xi_{dm} \; \alpha/2 \gg 1 $.

\medskip

The asymptotic expansion of the function $ b_{dm}(y) $ for large $ y $
follows expanding the integral representation eq.(\ref{dfb}) in inverse 
powers of $ \xi_{dm} \; y $. We obtain after calculation,
\be\label{asib}
b_{dm}(y)\buildrel{\xi_{dm} \; y \gg 1}\over= 3 + \frac52 \; \frac{I_4^{dm}}{(\xi_{dm} \; y)^2}
+ {\cal O}\left[\frac1{(\xi_{dm} \; y)^4}\right]\; .
\ee
where we used eq.(\ref{dfIn}).

\medskip

In the MD era we can approximate the inhomogeneous term $ a(y,\alpha) $ in eq.(\ref{volmd}) as
\be\label{defcmd}
a(y,\alpha)= a^{MD}(y,\alpha) \equiv
\frac{2 \, \xi_{dm} \; y}{\alpha} \int_0^{\infty}  \frac{Q \; dQ}{l^{NR}(y,Q)}
\left[ f_0^{dm}(Q) \; {\bar c}_{dm}^0(Q)+{\bar \phi}(0) \; \frac{df_0^{dm}}{d\ln Q}\right]
\; \sin \left[\frac{\alpha}2 \, Q \, l^{NR}(y,Q) \right] \; .
\ee
In eq.(\ref{volmd}) we can approximate the kernel $ N_\alpha(y,y') $ for $ y' > 0.1 $ 
according to eq.(\ref{Nnr}) and change the integration variable from
$ y' $ to $ s' $, defined as in eq.(\ref{defsy}), 
$$
s' \equiv -  {\rm Arg \, Sinh}\left( \frac1{\sqrt{y'}} \right) 
 \quad , \quad y' = y(s') = \frac1{\sinh^2 s'}  \quad , \quad 
\frac{ds'}{dy'} = \frac1{2 \; y' \; \sqrt{1+y'}} \quad .
$$
Then eq.(\ref{volmd}) becomes 
\be\label{cmd}
\frac{{\bar \Delta}_{dm}(y,\alpha)}{y} = g(y,\alpha) + \frac6{\alpha} 
\int_{s(1)}^{s(y)} ds' \; \Pi\left[\alpha\left(s(y)-s'\right)\right] \; {\bar \Delta}_{dm}(y(s'),\alpha) 
\ee
where $ \Pi(x) $ is given by eq.(\ref{defpi}), we used eq.(\ref{potmd}) and
\be\label{evf}
 g(y,\alpha) \equiv \frac{ a^{MD}(y,\alpha)}{y} + \frac{\kappa}{y} \; 
\; \int_0^1 \frac{dy'}{\sqrt{1+y'}} \; \left[N_\alpha(y,y')\; 
{\bar \phi}(y',\alpha)+N_\alpha^{\sigma}(y,y') \; {\bar \sigma}(y',\alpha)\right] \; .
\ee
The term proportional to $ {\bar \phi}(y,\alpha) $ in eq.(\ref{volmd}) becomes negligible
using eqs.(\ref{potmd}) and (\ref{asib}):
$$
\xi_{dm} \; b_{dm}(y) \; {\bar \phi}(y,\alpha) \buildrel{y > 1}\over= - 
\frac92 \; \frac{{\bar \Delta}_{dm}(y,\alpha)}{(\kappa \; y)^2} \; .
$$
Notice that $ a^{MD}(y,\alpha) $ is explicitly known. Once the Volterra 
equations (\ref{final})-(\ref{fsigma})
are solved from $ y = 0 $ till $ y = 1 $ we explicitly know $ g(y,\alpha) $ from
eq.(\ref{evf}). Then, the Volterra equation (\ref{cmd}) can be solved to find 
$ {\bar \Delta}_{dm}(y,\alpha) $ for $ y > 1 $.

\medskip

By setting
\be \label{dD2}
{\bar \Delta}_{dm}(y,\alpha) = y \; d_{dm}(y,\alpha)  \; ,
%\quad , \quad {\rm and ~ so} \quad \delta(y=1,\alpha)= 1 \; .
\ee
we obtain from the Volterra equation (\ref{cmd}) the Gilbert-type equation for the density fluctuations
valid when the DM particles are nonrelativistic. Notice that contrary to the original
Gilbert equation only valid in the MD dominated era, our equation is valid for all $ y \gtrsim 0.01 , \;
 z \lesssim 300000 $ well inside the RD era,
\be\label{dnmd}
d_{dm}(y,\alpha)= g(y,\alpha) + \frac6{\alpha} \int_{s(1)}^{s(y)} \frac{ds'}{\sinh^2 s'} \; 
\Pi\left[\alpha\left(s(y)-s'\right)\right] \; d_{dm}(y(s'),\alpha) \; .
\ee
where $ g(y,\alpha) $ is given by eqs.(\ref{defcmd}) and (\ref{evf}). 

\medskip

The memory piece in eq.(\ref{evf}) turns to be smaller than the first term $ a^{MD}(y,\alpha)/y $
by an order of magnitude or so as shown by numerical calculations. In addition,
for $ y > 1 $ (MD era) we can neglect the logarithmic dependence in $ Q $ present in $ l^{NR}(y,Q) $
eq.(\ref{lnr}). We therefore have for $ a^{MD}(y,\alpha) $ in the MD era from eq.(\ref{defcmd})
$$
a^{MD}(y,\alpha) =
\frac{\xi_{dm} \; y}{z(y)} \int_0^{\infty}  Q \; dQ \; 
\left[ f_0^{dm}(Q) \; {\bar c}_{dm}^0(Q)+{\bar \phi}(0) \; \frac{df_0^{dm}}{d\ln Q}\right]
\; \sin[z(y) \; Q] \; ,
$$
where $ s(y) $ is defined by eq.(\ref{defsy}) and 
$$ 
z(y) \equiv \alpha \left[s(y)+ \frac12 \; \log \left( 8 \; 
\xi_{dm} \right) \right] \; .
$$
For thermal and Gilbert initial conditions eqs.(\ref{condini}) we can 
express $ a^{MD}(y,\alpha) $ in terms
of the kernel $ \Pi(z) $ defined by eq.(\ref{defpi}) as
\be
\frac{a^{MD}(y,\alpha)}{y} = - \xi_{dm} \; \left\{ \left[{\bar \phi}(0) + \frac12 \right] \; \; 
\left[\frac{2 \, \Pi(z(y))}{z(y)} + \Pi'(z(y))\right] + j \; 
\left[\frac{\Pi(z(y))}{z(y)} - \frac12 \; \Pi'(z(y))\right]\right\} \; .
\ee
where $ j = 0 $ for thermal initial conditions and  $ j = 1 $ for Gilbert initial conditions.

\medskip

Eqs.(\ref{cmd}) and (\ref{dnmd}) have the same kernel as the Gilbert equation 
in the DM era \cite{gil,bdvs}
as it must be. The inhomogeneous term $ g(y,\alpha) $ differs to those in 
refs. \cite{gil,bdvs}: this is so because $ g(y,\alpha) $ 
takes into account the memory from the previous evolution 
of the fluctuations since the DM decoupling in the RD era considered here. 
In Appendix \ref{apef} we derive a Gilbert-type equation in the MD era
from eqs.(\ref{defcmd}), (\ref{evf}) and (\ref{dnmd}) valid in the MD era. 
The obtained Gilbert-type equation
eq.(\ref{voltnr}) contains an inhomogeneous term corresponding to temperature 
perturbation initial conditions plus a memory term including the contributions from the RD era.

\subsection{The Gilbert equation from the Volterra equation in the MD era.}\label{apef}

Eq.(\ref{dnmd}) can be written as
\bea\label{voltnr}
&& d_{dm}(v,\alpha) -  \frac6{\alpha} \int_{v_1}^{v} dv' \; y(v') \; 
\Pi\left[\alpha\left(v-v'\right)\right] \; d_{dm}(v',\alpha)= 
\xi_{dm} \; \left\{ \left[2 \, {\bar \phi}(0) + 1 \right] \; \; 
\left[\frac{\Pi\left[\alpha \; \left(v + v_0\right)\right]}{\alpha \; (v + v_0)} + 
\frac12 \; \Pi'\left[\alpha \; \left(v + v_0\right)\right] \right]+ \right. \cr \cr  
&& \left. 
+ g \; \left[ \frac{\Pi\left[\alpha \; \left(v + v_0\right)\right]}{\alpha \; (v + v_0)} - 
\frac12 \; \Pi'\left[\alpha \; \left(v + v_0\right)\right]\right] \right\}
= M[y(v),\alpha] \; ,
\eea
where $ g = 0 $ for thermal initial conditions (TIC) and $ g = 1 $ for 
Gilbert initial conditions (GIC),
\be \label{v0}
v \equiv s +  1 = 1 -  {\rm Arg \, Sinh}\left( \frac1{\sqrt{y'}} \right) 
\quad , \quad  v_0 \equiv \frac12 \; \log(8 \, \xi_{dm}) - 
1 \simeq 4.288 \ldots
+  \frac12 \ln \left( \frac{m}{\rm keV}\right)  +  
\frac16 \ln\left(\frac{g_d}{100}\right) \; , 
\ee
$ v_1 = 1 + s(1) $ and $ M[y,\alpha] $ stands for the memory term containing the contribution 
of the gravitational potential and anisotropic stress from the RD era
$$
M[y,\alpha] \equiv \frac{\kappa}{y} \; 
\; \int_0^1 \frac{dy'}{\sqrt{1+y'}} \; \left[N_\alpha(y,y')\; 
{\bar \phi}(y',\alpha)+N_\alpha^{\sigma}(y,y') \; 
{\bar \sigma}(y',\alpha)\right] \; .
$$

\medskip

On the other hand, the Gilbert equation in the MD era can be written as \cite{gil,bdvs}
\be\label{Gmd}
\delta_{MD}(y,\alpha) -  \frac6{\alpha} 
\int_0^{v(y)} dv' \; y_{MD}(v') \; \Pi\left[\alpha\left(v(y)-v'\right)\right] \; 
\delta_{MD}\left[y(v'),\alpha\right] = I_L[\alpha \; v(y)] \quad , 
\ee
where L = G or T. In the notation of ref. \cite{bdvs}  L = G or T corresponds
to Gilbert or temperature perturbation initial conditions, respectively,
\be\label{igit}
I_G(z) =  \frac{\Pi(z)}{z} \quad , \quad I_T(z) =  \frac13 \left[\Pi'(z) +
2 \; \frac{\Pi(z)}{z} \right] \quad  
y_{MD}(v) = \frac1{(1-v)^2} \quad {\rm and}  \quad
v_{MD}(y) = 1 -  \frac1{\sqrt{y}} \quad .
\ee
We see comparing eqs.(\ref{voltnr}) and (\ref{Gmd}) that the inhomogeneous terms
are different. The inhomogeneities in the Gilbert equation (\ref{Gmd}) contain the
functions $ I_G(\alpha \; v ) $ or  $ I_T(\alpha \; v )  $ while the inhomogeneities in the
Volterra equation (\ref{voltnr}) contain $ I_T[\alpha \; (v + v_0)] $ for TIC and a linear combination
of $ I_T[\alpha \; (v + v_0)] $ and $ I_G[\alpha \; (v + v_0)] $ for GIC.
Namely, the argument $ v $ in the inhomogeneous terms containing the kernels $ \Pi $ and $ \Pi' $ 
is shifted by the quantity $ v_0 $ given by eq.(\ref{v0}) 
(A similar shift was noticed in ref. \cite{bwu}). In addition, the inhomogeneous term $ M[y,\alpha] $,
memory of the RD era in the Volterra equation is necessarily absent in the  Gilbert equation
which only takes into account the MD era.

\medskip

In summary, choosing TIC at decoupling in the Volterra system of equations 
yields TIC at $ y=1 $ for the Gilbert equation. On the contrary, choosing GIC
at decoupling in the Volterra system of equations yields a linear combination
of TIC and GIC at $ y=1 $ for the Gilbert equation. One can thus say that TIC
are {\bf stable} under the evolution of the fluctuations.

In order to complete the comparison of eq.(\ref{voltnr}) in the late MD era
with the Gilbert equation (\ref{Gmd}), notice that
$$
v = s + 1 =  1 -{\rm Arg \, Sinh}\left( \frac1{\sqrt{y}} \right)  \buildrel{y \gg 1}\over=
1  -  \frac1{\sqrt{y}} \quad {\rm and ~hence}  \quad 
y(v) \buildrel{y \gg 1}\over= \frac1{\left(1-v\right)^2} \quad {\rm as ~ in ~ eq.(\ref{igit})}.
$$
We conclude that the Volterra integral equation (\ref{voltnr}) in the MD era 
is very close although not identical to the Gilbert equation in the MD era eq.(\ref{Gmd}).
The inhomogeneity in the Volterra integral equation (\ref{voltnr}) for TIC
has a factor in front and its argument is shifted  by the constant $ v_0 $
with respect to the usual inhomogeneity in the Gilbert equation (\ref{Gmd}).
For GIC a linear combination of Gilbert and thermal initial conditions appear at $ y=1 $ for the Gilbert 
equation. In addition, the term $ M[y,\alpha] $ containing the memory from the RD era is present
in the Volterra integral equation (\ref{voltnr}) while such term is absent in the Gilbert equation (\ref{Gmd}).

\section{Solving the Volterra equation for the DM density fluctuations (without anisotropic stress)}\label{numsol}

The formulation of the cosmological fluctuations evolution in terms of Volterra equations provides an efficient computational framework for both analytic and numerical treatment. In the following subsections we solve the Volterra equation for DM fluctuations in the absence of neutrinos, i. e. without anisotropic stress. First, we solve
the Volterra equation numerically for a wide range of wavenumbers. Second, we find the analytic solution at zero wavenumber.

\subsection{Numerical solution of the Volterra equation for a wide range of wavenumbers}
 
In the absence of anisotropic stress the radiation fluctuations during the RD era
can be treated in the fluid approximation. Neglecting the DM gravitational potential 
in the RD era, the gravitational potential is given in the fluid approximation
by (see ref. \citep{dod} and Appendix \ref{potr})
\be\label{firad}
{\breve \phi}(y,\alpha) = {\bar \phi}(y,\alpha) = {\breve \psi}(y,\alpha) = 3 \; 
\frac{\sqrt3}{\kappa \; y} \; j_1\left(\frac{\kappa \; y}{\sqrt3}\right) \quad , \quad
{\breve \phi}(0,\alpha) = 1 \; ,
\ee
%$ \psi(0,k) $ is the primordial value of the photon gravitational potential eq.(\ref{fikp}),
$ j_1(x) $ is the spherical Bessel function of order one.
Notice that $ {\displaystyle \lim_{x \rightarrow 0}} \; j_1(x)/x = 1/3 \; .$

In the MD era we can neglect the gravitational potential produced by the radiation and
take as gravitational potential the one sourced by the DM fluctuations eq.(\ref{fidm}).
In the absence of anisotropic stress, the DM fluctuations $ {\bar \Delta}_{dm}(y,\alpha) $
from eq.(\ref{volmd}) obeys the Volterra equation:
\be\label{sinneu}
{\bar \Delta}_{dm}(y,\alpha)=  a(y,\alpha) + y \; \xi_{dm} \; b_{dm}(y) \; {\breve \phi}(y,\alpha)
 + \kappa \; \int_0^y \frac{dy'}{\sqrt{1+y'}} \; N_\alpha(y,y')\; {\breve \phi}(y',\alpha)  \; ,
\ee
where $ a(y,\alpha) $ is given by eq.(\ref{asd3}) with $ {\bar \phi}(0) = 1 $.

We present here the numerical solution of eq.(\ref{sinneu})
where we smoothly match the gravitational potentials given by eqs.(\ref{firad})
and (\ref{fidm}).  The full numerical analysis of the system of Volterra
equations (\ref{final})-(\ref{fsigma}) including the
anisotropic stress will be the subject of future work where we will also compare 
our approach with the numerical solution of the ODE hierarchy of B-V 
equations \cite{mab}-\cite{sz}.

\medskip

For $ y \gtrsim 0.01 $ the DM particles become nonrelativistic eq.(\ref{sinneu}) simplifies
and takes the form of eq.(\ref{dnmd})
\be\label{snnr}
d_{dm}(y,\alpha)= h(y,\alpha) + \frac6{\alpha} \int_{s(1)}^{s(y)} \frac{ds'}{\sinh^2 s'} \; 
\Pi\left[\alpha\left(s(y)-s'\right)\right] \; d_{dm}(y(s'),\alpha) \quad , \quad 
d(y,\alpha) \equiv \frac{{\bar \Delta}_{dm}(y,\alpha)}{y} \; ,
\ee
where  $ l^{NR}(y,Q) $ is defined by eq.(\ref{lnr}),
\be 
h(y,\alpha) \equiv  \frac{ a^{MD}(y,\alpha)}{y} + \frac{\kappa}{y} \; 
\; \int_0^1 \frac{dy'}{\sqrt{1+y'}} \; N_\alpha(y,y')\; {\breve \phi}(y',\alpha)
\ee
and we have neglected the memory piece from the DM fluctuations in the UR regime but kept
the gravitational potential of the photons which is dominant.
Eq.(\ref{snnr}) is a closed integral equation of Volterra type that
determines approximately $ d(y,\alpha) $. We have checked numerically that
eq.(\ref{snnr}) reproduces the solutions of the full  Volterra equation (\ref{sinneu}) within
a few percent. 

\medskip

From the numerical resolution of the Volterra equation (\ref{sinneu}) we find the 
normalized density contrast 
\be \label{delsom}
{\breve\delta}(y,\alpha) \equiv \frac{\delta(y,\alpha)}{\delta(0,\alpha)}=
-\frac1{2 \, I_3^{dm}} \; 
\frac{{\bar\Delta}_{dm}(y,\alpha)}{y+1} \quad , \quad 
\delta(0,\alpha) = -\frac{2 \, I_3^{dm}}{\xi_{dm}} \quad , \quad 
{\breve\delta}(0,\alpha) = 1 \quad .
\ee
The density contrast $ \delta(y,\alpha) $ is given by eq.(\ref{Ddi}) and 
eq.(4.36) in ref. \cite{uno}.

\medskip

\begin{figure}[h]
\begin{center}
\begin{turn}{-90}
\psfrag{"ldca0.dat"}{$\log|{\breve{\Delta}}_{dm}(y,0)| $ vs. $ \log_{10} \; y $}
\psfrag{"ldca0003.dat"}{$\log| {\breve\delta}(y,\alpha=0.003)| $ vs. $ \log_{10} \; y $}
\psfrag{"ldca0005.dat"}{$\log| {\breve\delta}(y,\alpha=0.005)| $ vs. $ \log_{10} \; y $}
\psfrag{"ldca001b.dat"}{$ \log|{\breve\delta}(y,\alpha=0.01)| $ vs. $ \log_{10} \; y $}
\psfrag{"ldca003.dat"}{$ \log|{\breve\delta}(y,\alpha=0.03)| $ vs. $ \log_{10} \; y $}
\psfrag{"ldca01b.dat"}{$ \log|{\breve\delta}(y,\alpha=0.1)| $ vs. $ \log_{10} \; y $}
\includegraphics[height=16.cm,width=10.cm]{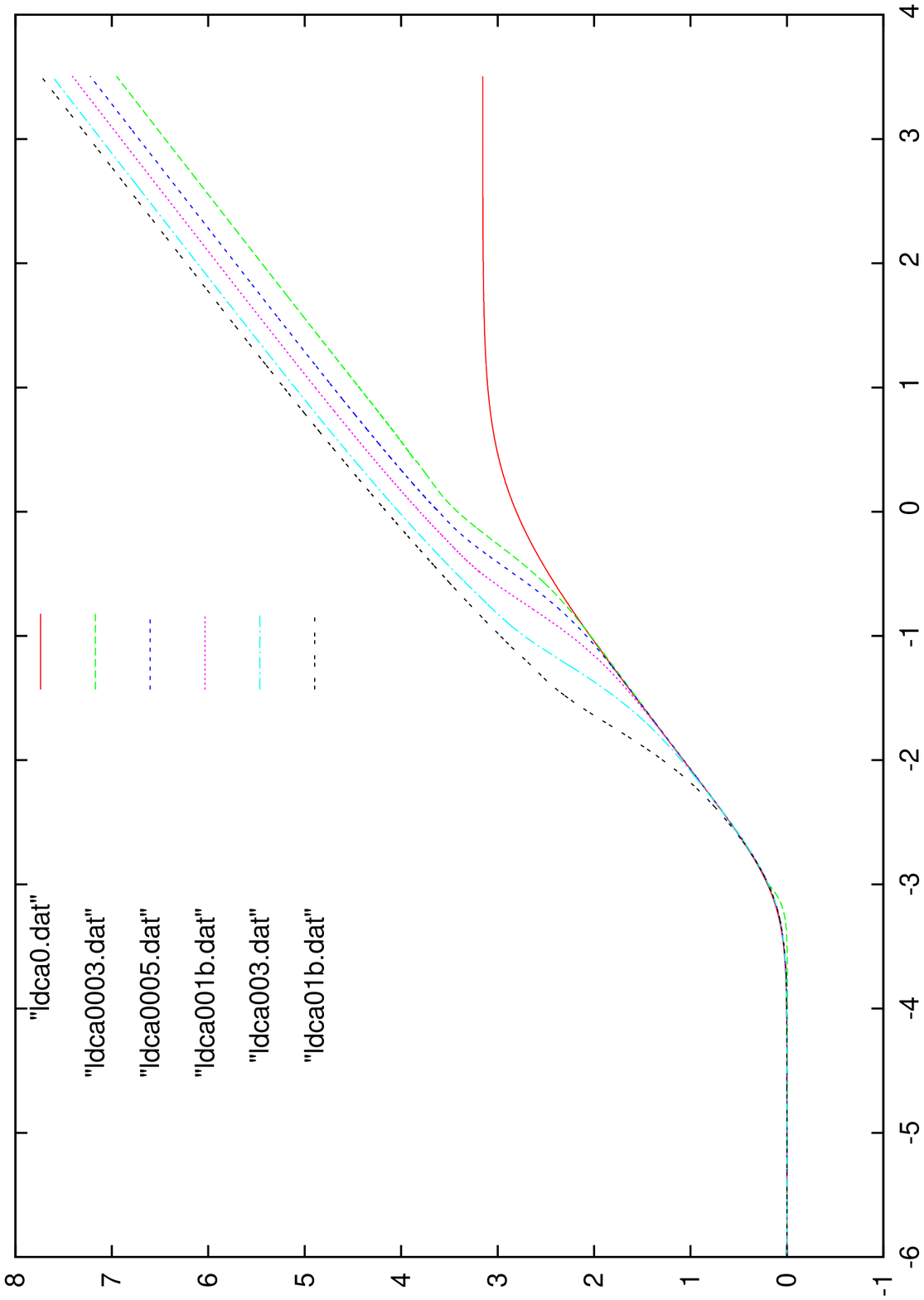}
\psfrag{"lda01b.dat"}{$ \log|{\breve\delta}(y,\alpha=0.1)| $ vs. $ \log_{10} \; y $}
\psfrag{"lda05.dat"}{$ \log|{\breve\delta}(y,\alpha=0.5)| $ vs. $ \log_{10} \; y $}
\psfrag{"lda07.dat"}{$ \log|{\breve\delta}(y,\alpha=0.7)| $ vs. $ \log_{10} \; y $}
\psfrag{"lda1.dat"}{$ \log|{\breve\delta}(y,\alpha=1.)| $ vs. $ \log_{10} \; y $}
\psfrag{"ldca3.dat"}{$ \log|{\breve\delta}(y,\alpha=3.)| $ vs. $ \log_{10} \; y $}
\psfrag{"ldca10.dat"}{$ \log|{\breve\delta}(y,\alpha=10.)| $ vs. $ \log_{10} \; y $}
\includegraphics[height=16.cm,width=10.cm]{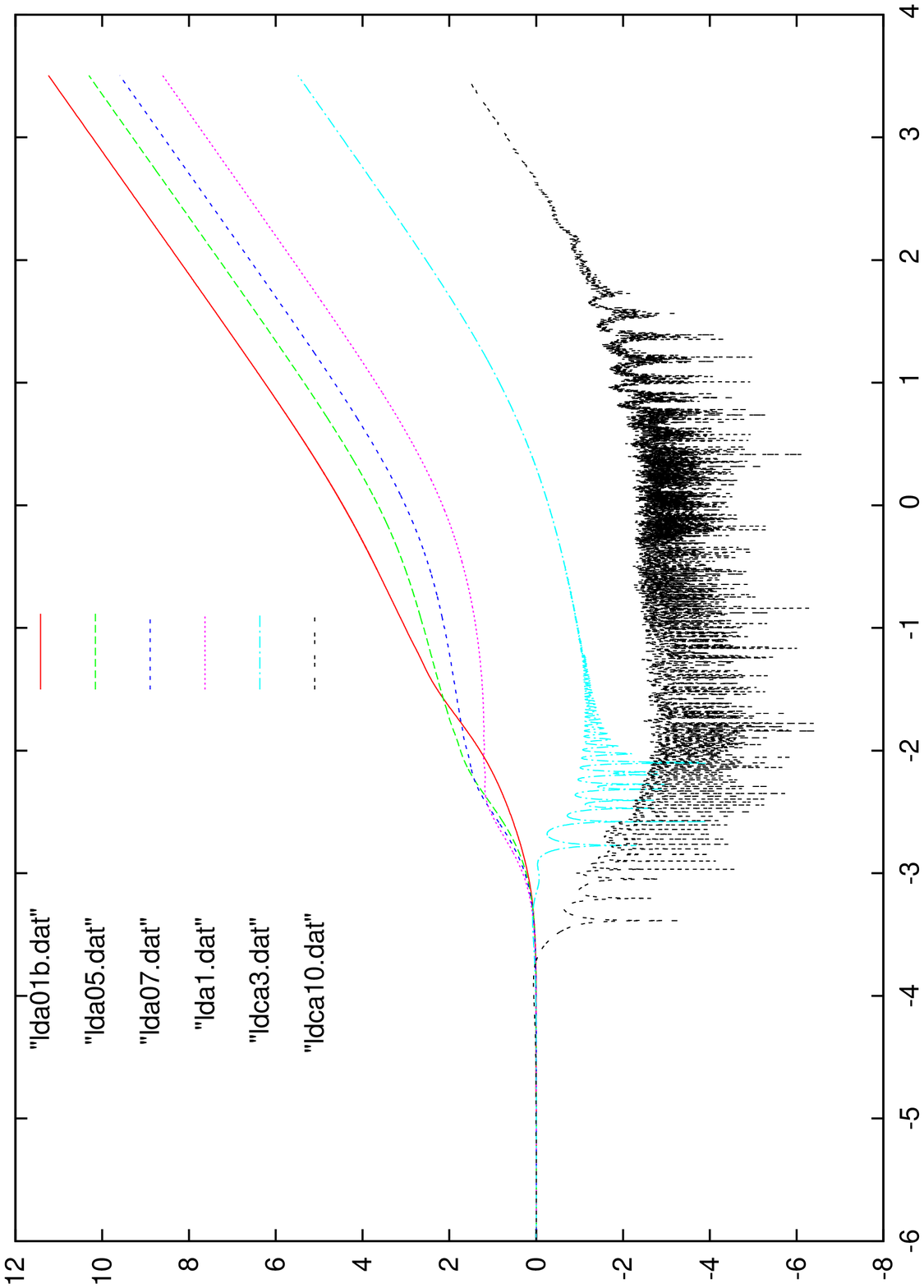}
\end{turn}
\caption{The ordinary logarithm of the normalized density contrast
$ {\breve\delta}(y,\alpha) $ vs. $ \log_{10} y $
from the numerical resolution of the  Volterra equation
(\ref{sinneu}) for DM fermions in thermal equilibrium with $ m = 0.6736 $ keV
and for sterile neutrinos in the DW model with $ m = 1.685 $ keV.
(Both models yield identical density fluctuations for a given value of 
$ \xi_{dm} $). There is a value $ \alpha = \alpha_c \simeq 0.1 $ 
determining the transition between two regimes. We separately display the 
plots corresponding to $ \alpha < \alpha_c $ and $ \alpha > \alpha_c $. 
We see that for $ \alpha < \alpha_c $ and fixed $ y , \;
 {\breve\delta}(y,\alpha) $ {\bf increases} for increasing $ \alpha $ while
the opposite happens for $ \alpha > \alpha_c $. 
$ {\breve\delta}(y,\alpha=0) $ is plotted from the analytic solution 
eq.(\ref{2condi})-(\ref{fikcero}) 
for TIC. We see that the different curves have essentially the same shape and
are shifted from each other in an almost selfsimilar manner indicating that
$ {\breve\delta}(y,\alpha) $ is mainly a function of $ \kappa \; y $.
$ {\breve\delta}(y,\alpha) $ generically grows with $ y $ for fixed 
$ \alpha < 1 $ while it exhibits oscillations starting in the RD era 
for $ \alpha > 1 $ which become stronger as $ \alpha $ grows. 
$ {\breve\delta}(y,\alpha) $ becomes proportional to $ y $ at 
sufficiently late times. The larger is $ \alpha $, the later starts 
$ \delta(y,\alpha) $ to grow proportional to $ y $ and 
the later the oscillations remain.} 
\label{flucd}
\end{center}
\end{figure}

We depict in fig. \ref{flucd} the logarithm of the absolute value of the 
normalized density contrast for fermions with $ \xi_{dm} = 5000 $ 
which corresponds to DM fermions in thermal equilibrium with $ m = 0.6736 $ keV
and sterile neutrinos in the DW model with $ m = 1.685 $ keV
(both models yield identical density fluctuations for a given value of 
$ \xi_{dm} $). In both cases we used thermal initial conditions. 

\medskip

The density contrast generically grows with $ y $ for fixed $ \alpha < 1 $ 
while it exhibits oscillations starting in the RD era 
for $ \alpha > 1 $ which become stronger as $ \alpha $ grows 
(see fig. \ref{flucd}). As expected, the Jeans' unstability makes the 
density contrast proportional to $ y $ (to the scale factor) at 
sufficiently late times. The larger is $ \alpha $, the later starts 
$ \delta(y,\alpha) $ to grow proportional to $ y $ (see fig. \ref{flucd}). 
Also, the larger is $ \alpha > 1 $, the later the oscillations remain.

\medskip

There exists a  value $ \alpha = \alpha_c \simeq 0.1 $ determining the transition between
two regimes. 
We separately display the plots corresponding to $ \alpha < \alpha_c $
and $ \alpha > \alpha_c $. We find that for $ \alpha < \alpha_c $ and fixed $ y , \;
{\breve\delta}(y,\alpha) $ {\bf increases} for increasing $ \alpha $ while
the opposite happens for $ \alpha > \alpha_c $. Namely,  $ {\breve\delta}(y,\alpha) $
at fixed $ y $ {\bf decreases} for increasing $ \alpha $. We see from fig. \ref{flucd}
that the curves for $  \log_{10} |{\breve\delta}(y,\alpha)| $ vs. $ \log_{10} \; y $ keep
bending for decreasing $ \alpha \to 0 $ towards the $ \alpha = 0 $ curve.
[The $ \alpha = 0 $ curve is obtained analytically in eqs.(\ref{2condi}) and (\ref{fikcero}) below]. 

\medskip

In figs. \ref{flucd} we see that for both $ \alpha < \alpha_c $ and $ \alpha > \alpha_c $
varying  $ \alpha $ shifts the curves $ {\breve\delta}(y,\alpha) $
vs. $ y $ with respect to each other but keeping their form essentially unchanged.
This property indicates that $ {\breve\delta}(y,\alpha) $
mainly depends on $ \alpha $ and $ y $ through the product $ \alpha \; y $, namely
in a selfsimilar manner. 

\medskip

We have computed $ {\breve\delta}(y,\alpha) $ in the $ \chi $-model for 
sterile neutrinos and found curves quite similar to the thermal case 
fig. \ref{flucd}.

\subsection{Analytic solution of the Volterra equation at zero wavenumber}\label{solk0}

At $ \alpha = 0 $ (that is, $ k = 0 $), the Volterra equation (\ref{sinneu}) (zero
anisotropic stress) can be solved in close form since from eq. (\ref{nucL2})
its kernel vanishes: $ N_{\alpha=0}(y,y') = 0 $.

\medskip

The inhomogeneous term  $ a(y,0) $ in the Volterra equation (\ref{sinneu}) 
becomes using eq.(\ref{asd3}) 
\be \label{acero}
a(y,0)=- \, y \; \xi_{dm} \; b_{dm}(y) + \int_0^{\infty}  Q^2 \; dQ \; \varepsilon(y,Q) \; 
f_0^{dm}(Q) \; {\bar c}_{dm}^0(Q) \; ,
\ee
Therefore, the Volterra equation (\ref{sinneu}) at $ \alpha = 0 $
simply relates the DM density fluctuations in terms of the gravitational potential as
\be\label{alf0}
{\bar \Delta}_{dm}(y,0)= \int_0^{\infty}  Q^2 \; dQ \; \varepsilon(y,Q) \; 
f_0^{dm}(Q) \; {\bar c}_{dm}^0(Q) + 
y \; \xi_{dm} \; b_{dm}(y) \left[{\breve \phi}(y,0) - 1 \right] \; .
\ee
More explicitly, for thermal (TIC) and Gilbert (GIC) initial conditions eq.(\ref{condini}),
$ {\bar \Delta}_{dm}(y,0) $ takes the form
\be\label{2condi}
{\bar \Delta}_{dm}(y,0)=\left\{\begin{array}{l} 
y \; \xi_{dm} \; b_{dm}(y) \left[{\breve \phi}(y,0) -  \frac32 \right] \quad {\rm for ~ TIC} \quad , \\ \\
-2 \; \int_0^{\infty}  Q^2 \; dQ \; \varepsilon(y,Q) \; f_0^{dm}(Q)+ 
y \; \xi_{dm} \; b_{dm}(y) \left[{\breve \phi}(y,0) - 1 \right] \quad {\rm for ~ GIC} \quad .
\end{array} \right.
\ee
We can obtain $ {\breve \phi}(y,0) $ solving the hydrodynamic equations for the radiation fluctuations
(\ref{hidro2})-(\ref{hidro}) together with the linearized Einstein equations for the gravitational 
potential in the $ k \to 0 $ limit
\bea\label{eilif}
&&-k^2 \; \phi_r(\eta,\vk) = 16 \, \pi \; G \; a^2(\eta) \; \rho_\gamma(\eta)
\left[\Theta_{r,0}(\eta,\vk) + \frac3{k} \; h(\eta) \;  \Theta_{r,1}(\eta,\vk)
\right] \; , \\ \cr
&&3 \; h(\eta)  \; \frac{\partial \phi}{\partial \eta} + 
k^2 \; \phi(\eta, \vk) + 3 \; h^2(\eta) \; \psi(\eta, \vk) = -4 \, 
\pi \; G \; a^2(\eta) \; \left[4 \, \rho_r(\eta) \; \Theta_{r,0} (\eta, \vk)
+  \rho_{dm}(\eta) \; \delta_{dm} (\eta, \vk)\right]\; ,\label{eilif2}
%\frac6{\eta^2[1 + \eta/(4 \; \eta^*)]^2} 
\eea
where $ \phi_r(\eta,\vk) $ stands for the radiation contribution to the 
gravitational potential and $ \rho_{dm}(\eta) \; \delta_{dm} (\eta, \vk) $ for the DM fluctuations.
Eq.(\ref{eilif2}) can be written in dimensionless variables as
\be\label{eilif3}
y \; \left[1+ {\cal R}_0(y)\right] \; \frac{d{\bar \phi}}{dy} + \frac13 \left(\kappa \; y \right)^2 
{\bar \phi}(y,\alpha) 
+ \left[1 +  {\cal R}_0(y) \right] \; {\breve \psi}(y,\alpha)  = - 2 \; {\bar \Theta}_{r,0}(y,\alpha)
- \frac12 \;  {\cal R}_0(y) \;  {\bar\delta}_{dm}(y,\alpha) \; ,
\ee
where $ {\cal R}_0(y) $ is defined in eqs.(\ref{defR})-(\ref{R}) and we used eq.(\ref{fotones}).

\medskip

Since the left hand side of eq.(\ref{eilif}) vanishes at $ k = 0 $ we have 
\be\label{teta1}
\Theta_{r,1}(\eta,\vk) \buildrel{k \to 0}\over= - \frac{k}{3 \; h(\eta)}
\; \Theta_{r,0}(\eta,\vk) + {\cal O}(k^3) \quad .
\ee
We neglect radiation momenta higher than $ l = 1 $ thus 
neglecting the anisotropic stress and set 
$ \psi(\eta, \vk) = \phi(\eta, \vk) $.  

We have from eq.(\ref{hidro2}) in the $ k \to 0 $ limit
\be\label{tetfi0}
\Theta_{r,0}(\eta,0) - \phi(\eta,0) = c
\ee
where $ c $ is a constant. 

\medskip

The initial values $ \psi(0, \vk) $ and $ \Theta_{r,0}(0,\vk) $
are related by the $ \eta \to 0 $ limit of eq.(\ref{eilif2}) as
\be\label{tetfii}
\psi(0, \vk) = - 2 \; \Theta_{r,0}(0,\vk) \; ,
\ee
up to small $ 1/\xi_{dm} $ corrections. 
We thus find from eqs.(\ref{tetfi0}) and (\ref{tetfii}),
\be\label{cfite}
c = -\frac32 \; \phi(0,0) = 3 \; \Theta_{r,0}(0,0) \; .
\ee
Inserting eqs.(\ref{teta1}) and (\ref{tetfi0}) in eq.(\ref{hidro})
yields
$$
\frac{d}{d \eta}\left[\frac{\Theta_{r,0}}{h(\eta)}\right]+ 
2 \;\Theta_{r,0}(\eta,0) = c \; ,
$$
which in terms of the variable $ y $ becomes
$$
y \; \frac{d\Theta_{r,0}}{dy} + \left[3 - \frac{y}{2 \; [{\cal R}_0(y)+1]} \; \frac{d{\cal R}_0(y)}{dy}
\right] \; \Theta_{r,0}(y)=c \; .
$$
This first order differential equation can be resolved
with the explicit solution
\be
\Theta_{r,0}(y) = \Theta_{r,0}(0) \; \frac2{5 \; y^3}
\left[3 \; y^3 - y^2 + 4 \, y + 8 \left(1 - \sqrt{y+1}\right)\right] \; ,
\ee
up to small corrections of the order $ 1/\xi_{dm} $ because we set here $ {\cal R}_0(y) = y $ 
[see eqs.(\ref{defR})-(\ref{R})].

Then, the gravitational potential follows from eqs.(\ref{tetfi0})
and (\ref{cfite}) and we recover the known expression \cite{dod}
\be\label{fikcero}
{\breve \phi}(y,0)=\frac{\phi(y,0)}{\phi(0,0)}= \frac32-
\frac{\Theta_{r,0}(y)}{2 \; \Theta_{r,0}(0)}=\frac1{10 \; y^3}
\left[9 \; y^3 +2\, y^2 - 8 \, y + 16 \left(\sqrt{y+1}-1\right)\right] \; .
\ee
For zero or small redshift, eq.(\ref{fikcero}) becomes
\be\label{ficeroas}
{\breve \phi}(y,0)\buildrel{y \gg 1}\over=\frac9{10} + \frac1{5 \, y} + {\cal O}\left( \frac1{y^2} \right) \quad .
\ee
It must be noticed that the known expression eq.(\ref{fikcero}) for the superhorizon gravitational
potential (see for example ref. \cite{dod}) follows here {\bf solely} from the hydrodynamic 
equations for the radiation (\ref{hidro2})-(\ref{hidro}) combined with eq.(\ref{teta1}).
[Eq.(\ref{teta1}) follows from the first linearized Einstein equation
(\ref{eilif}) in the $ k \to 0 $ limit].
Namely, $ {\breve \phi}(y,0) $ and $ {\bar \Theta}_{r,0}(y,0) $ are obtained {\bf without} 
specifying the sources of the DM and radiation fluctuations.

\medskip

We can find the matter source of the superhorizon 
gravitational potential $ {\breve \phi}(y,0) $ by inserting eq.(\ref{fikcero}) in the left hand
side of eq.(\ref{eilif2}) for $ \psi(\eta, \vk) = \phi(\eta, \vk) $ and 
$ k = 0 $. We obtain using the dimensionless variable $ y $
\be\label{2ecfi0}
\left[1+{\cal R}_0(y)\right] \; \left[y \; \frac{d}{dy} + 1 \right]{\breve \phi}(y,0)=- \left[2  + 
2 \; {\cal R}_0(y) - \frac{y}{2} \; \frac{d{\cal R}_0(y)}{dy} \right]{\bar \Theta}_{r,0}(y,0) \; , 
\ee
Contrasting eq.(\ref{2ecfi0}) with eq.(\ref{eilif3}) and using eq.(\ref{R})
{\bf implies} a DM source 
\be\label{fnr}
{\bar \delta}_{dm} (y, 0) = \left(4 - \frac{d \ln {\cal R}_0}{d \ln y}\right)
\; {\bar \Theta}_{r,0} (y, 0)= \left\{\begin{array}{l} 
4 \; {\bar \Theta}_{r,0} (y, 0) \quad {\rm for} \quad 
\xi_{dm} \; y \lesssim 1 \quad , \\ \\
3 \; {\bar \Theta}_{r,0} (y, 0) \quad {\rm for} \quad 
\xi_{dm} \; y \gtrsim 1 \quad .
\end{array} \right.
\ee
That is, combining the linearized Einstein equations with the hydrodynamic
equations for the radiation {\bf requires} for consistency a precise 
{\bf relation} between the dark matter and radiation fluctuations.
This is a consequence of the fact that the Einstein equations constrain their sources as was first 
noticed in ref. \cite{eih} in a completely different context.

\medskip

Inserting eq.(\ref{ficeroas}) into eq.(\ref{2condi}) yields for the DM density fluctuations today
$$
{\bar \Delta}_{dm}(y,0)\buildrel{y \gg 1}\over= - \xi_{dm} \; y
\left\{\begin{array}{l} 
\displaystyle \frac95  - \frac3{5 \; y}\left[1 +\displaystyle {\cal O}\left( \frac1{y} \right)\right]
\quad {\rm for ~ TIC} \quad \quad , \\ \\
\displaystyle\frac{23}{10}  - \frac3{5 \; y} \; \left[1 + \displaystyle{\cal O}\left( \frac1{y} \right)\right]
\quad {\rm for ~ GIC} \quad .
\end{array} \right.
$$
The normalized density contrast eq.(\ref{delsom}) becomes today and for zero wavenumber,
\be\label{contra0}
{\breve\delta}(y,0)\buildrel{y \gg 1}\over=\frac{\xi_{dm}}{10 \; I_3^{dm}}
\left\{\begin{array}{l} 
\displaystyle 9  \left[1 +\displaystyle {\cal O}\left( \frac1{y} \right)\right]
\quad {\rm for ~ TIC} \quad \quad , \\ \\
\displaystyle\frac{23}2  \left[1 + \displaystyle{\cal O}\left( \frac1{y} \right)\right]
\quad {\rm for ~ GIC} \quad .
\end{array} \right.
\ee
We depict $ \log_{10} |{\breve{\delta}}(y,0)| $ vs. $ \log_{10} \; y $  in fig. \ref{flucd}
for TIC. Fig. \ref{flucd} exhibits this constant behaviour in 
$ {\breve{\delta}}(y,0)$ for large $ y $.

\begin{figure}
\begin{center}
\begin{turn}{-90}
\psfrag{"dhoyRgam.dat"}{Thermal fermions}
\psfrag{"trucho.dat"}{}
\psfrag{"edhoygam.dat"}{sterile neutrinos in the $ \chi $-model}
\includegraphics[height=16.cm,width=10.cm]{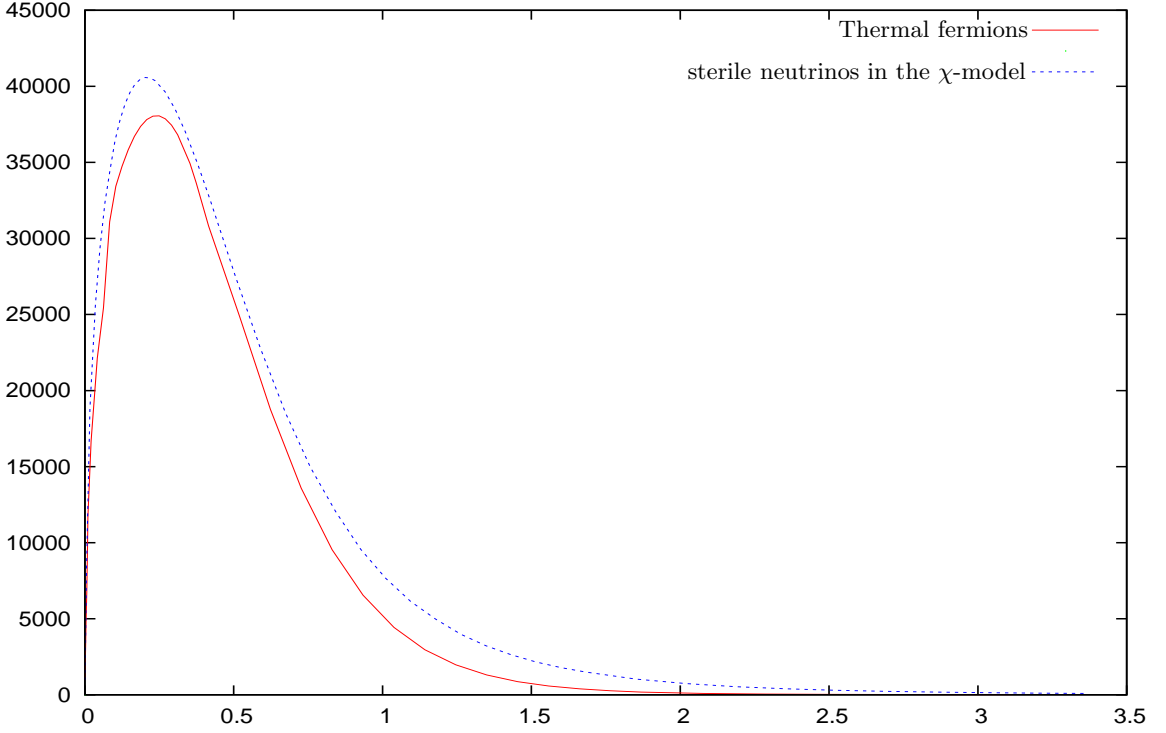}
\end{turn}
\caption{The transfer function today $ T(\gamma) $ vs. 
$ \gamma = \sqrt{\frac{I_4^{dm}}3} \; \alpha $ for  $ \xi_{dm} = 5000 $.
The (red) solid line curve is for DM fermions in thermal equilibrium with $ m = 0.6736 $ keV
and sterile neutrinos out of equilibrium in the DW model with $ m = 1.685 $ keV. The (blue) dotted 
line corresponds to sterile neutrinos out of equilibrium in the $ \chi $-model with 
$ m = 0.7203 \; \tau^{-1/4} $ keV where $ \tau $ is a coupling constant [see eq.(\ref{3mod})].
That is, $ 0.9365 < m/{\rm keV} < 1.665 $ [see eq.(\ref{3xi})].
$ \gamma $ is defined by eq.(\ref{dgam}). The presence here of a single maximum at 
$ \gamma = \gamma_c \simeq 0.2 $ is consistent with  the curves for $ {\breve\delta}(y,\alpha) $ 
and the value of $ \alpha_c $ in figs. \ref{flucd}. Notice that the two transfer functions turn out
to be very similar although they describe quite different dynamics.}
\label{talfa}
\end{center}
\end{figure}

\subsection{The transfer function for the density contrast}

The transfer function at redshift $ z $ can be defined as the density contrast at redshift 
$ z \geq 0 $ normalized
by its initial value and then normalized by the whole expression at $ k = 0 $. That is,
$$
T(y, \alpha) \equiv \frac{{\breve\delta}(y,\alpha)}{{\breve\delta}(y,0)} \quad , \quad T(y,0) = 1
\; .
$$
The transfer function today becomes
\be
T(\alpha) = {\displaystyle \lim_{y \gg 1}} T(y, \alpha) =\frac{10 \; I_3^{dm}}{9 \; \xi_{dm}} \;
{\breve\delta}(y,\alpha) \quad {\rm for ~ TIC} \; {\rm and} \quad
T(\alpha) =\frac{20 \; I_3^{dm}}{23 \; \xi_{dm}} \;
{\breve\delta}(y,\alpha) \quad {\rm for ~ GIC}
\quad , \quad T(0) = 1 \; ,
\ee
and we used eq.(\ref{contra0}).

\medskip

We plot in fig. \ref{talfa} the zero redshift transfer function $ T(\gamma) $
vs. $ \gamma $ for $ \xi_{dm} = 5000 $ and TIC. 
The (red) solid line curve is for DM fermions decoupling in thermal equilibrium
with $ m = 0.6736 $ keV and sterile neutrinos out of equilibrium in the DW model with $ m = 1.685 $ keV. 
The (blue) dotted line corresponds
to sterile neutrinos out of equilibrium in the $ \chi $-model with $ m = 0.7203 \; \tau^{-1/4} $ keV.
That is, $ 0.9365 < m/{\rm keV} < 1.665 $ [see eq.(\ref{3xi})].
The variable $ \gamma $ is defined as
\be\label{dgam}
\gamma \equiv \alpha \; \sqrt{\frac{I_4^{dm}}3} \; .
\ee
We find that  the transfer functions have a single maximum at $ \alpha_c $ consistent
with the behaviour of  $ {\breve\delta}(y,\alpha) $ in figs. \ref{flucd}.

\medskip

Notice that $ T(\gamma) $ grows fast with $ \gamma $  for $ 0 < \gamma < \gamma_c $
as we see from fig. \ref{talfa}. 
Previous calculations of the transfer function in refs. \cite{BBKS,turokWDM} 
and with better precision in ref. \cite{bdvs} only exhibit the 
portion of $ T(\gamma) $ where it decreases with $ \gamma $.

\medskip

The transfer function computed solely in the MD era monotonically
decreases with $ \gamma $ for growing $ \gamma \geq 0 $ \cite{dod,bdvs}. 
The {\bf new piece} of $ T(\gamma) $ increasing with $ \gamma $  for 
$ 0 < \gamma < \gamma_c $ comes from the behaviour of the DM fluctuations 
in the RD era computed here. $ \alpha_c $ and $ \gamma_c $ correspond here 
to a wavenumber $ k_c \simeq 1.6/$Mpc. 

\medskip

The transfer function for DM fermions decoupling in thermal equilibrium
and for sterile neutrinos out of equilibrium in the $ \chi $-model turn out
to be very similar as seen from fig. \ref{talfa}.

\section{Fermions in thermal equilibrium and sterile neutrinos
out of equilibrium}\label{neuest}

The sterile neutrino is a serious candidate for WDM \cite{dw,neus,modelos}.
The freezed out distribution of sterile neutrinos turns to be out of thermal
equilibrium in most models \cite{fuera}. We consider here two  sterile neutrino models
for illustration. The Dodelson-Widrow model (DW) \cite{dw} and the 
$ \chi $ model of ref. \cite{modelos}.

\medskip

The freezed-out DM distributions are given by
\be\label{3mod}
{\rm DW \; model} : \; \; f_0^{DW}(Q) =
\frac{f_0}{m} \frac1{e^{q/T}+1} \; , \; f_0 \simeq 0.043 \; {\rm keV}
\quad , \quad \chi \; {\rm model} : \; \;  f_0^{\chi}(Q) =\tau \; f_0^{dm}(Q) \; , \;
\quad 0.035 \lesssim \tau \lesssim 0.35 \; ,
\ee
where $ \tau $ is a coupling constant and the normalized DM distribution 
function for the  $ \chi $ model \cite{distest,dvs} takes the form 
\vskip -0.4 cm
\be \label{fnue}
f_0^{dm}(Q) = \frac4{3 \; \zeta(5) \; \sqrt{\pi \; Q}}
\sum_{n=1}^{\infty} \frac{e^{-n \, Q}}{n^{\frac52}} \; .
\ee
The normalized DM distribution in the DW model \cite{dw} is identical to
the normalized Fermi-Dirac distribution 
\be\label{fdsd}
f_0^{dm}(Q) = \frac2{3 \, \zeta(3)} \; \frac1{e^Q + 1} \; .
\ee
The simple formula eqs.(\ref{3mod})-(\ref{fdsd}) for the DW freezed-out distributions were 
given in \cite{dw} and are widely used in the literature. A more sophisticated freezed-out distribution
is derived in ref. \cite{als}. 

\begin{figure}[h]
\begin{center}
\begin{turn}{-90}
\psfrag{"dfd.dat"}{Fermi-Dirac distribution}
\psfrag{"dest.dat"}{ sterile neutrinos in the $ \chi $ model}
\includegraphics[height=13.cm,width=8.cm]{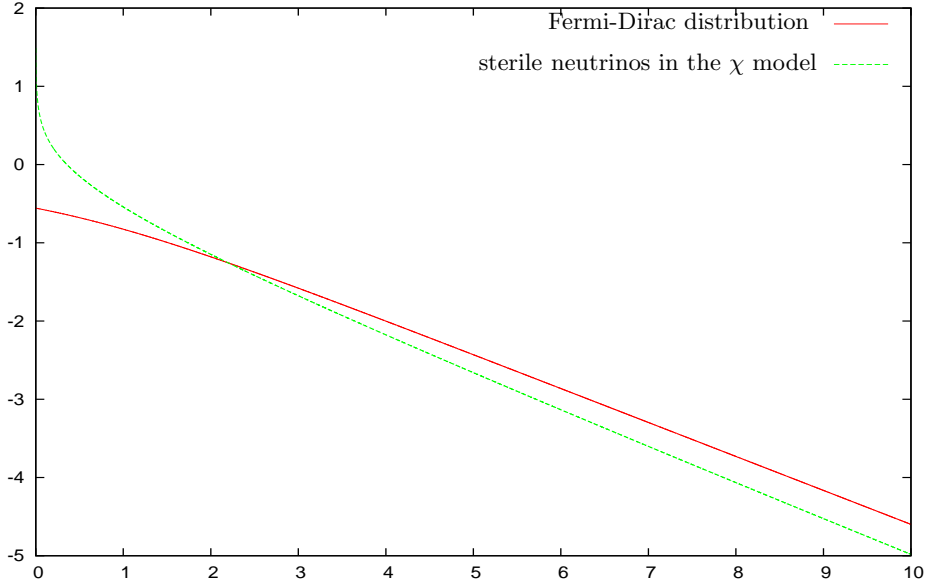}
\end{turn}
\caption{The ordinary logarithm of the normalized distribution functions $ f_0^{dm}(Q) $ vs. $ Q $
for fermions at thermal equilibrium (which is identical to the out of thermal equilibrium DW model)
and for sterile neutrinos out of thermal equilibrium in the $ \chi $ model.}
\label{distp}
\end{center}
\end{figure}

We plot in fig. \ref{distp} the normalized distribution functions $ f_0^{dm}(Q) $
for fermions in thermal equilibrium (which is identical to the DW model)
and for sterile neutrinos out of thermal equilibrium in the $ \chi $ model.

\medskip

We find from eqs.(\ref{forxi}) and (\ref{3mod}) the values for the parameters $ \xi_{dm} $ and $ N_{dm} $
in the three DM fermion models considered here: 
\bea\label{3xi}
&&\xi_{dm}^{FD} = 6721 \; (g_{dm})^\frac13 \; \left(\frac{m}{\rm keV}\right)^\frac43 \quad , \quad
\xi_{dm}^{DW} = 2355 \;  (g_{dm})^\frac13 \; \frac{m}{\rm keV} \quad , \quad
\xi_{dm}^{\chi} = 6146 \; (g_{dm} \; \tau)^\frac13 \; \left(\frac{m}{\rm keV}\right)^\frac43 \quad , 
\cr \cr \cr
&& N_{dm}^{FD} = 1.805 \quad , \quad N_{dm}^{DW} = 0.07765 \; \frac{\rm keV}{m} \quad , \quad
N_{dm}^{\chi} = 1.380 \;  \tau \; .
\eea
Notice that the  out of thermal equilibrium distribution is {\bf larger}
than the equilibrium distribution for small momenta $ Q \lesssim 2 $
while the opposite happens for  $ Q \gtrsim 2 $. This can be explained
by the general mechanism of thermalization: the momentum cascade towards the
ultraviolet \cite{ddv}.
The distributions out of equilibrium therefore display larger occupation at 
low momenta and smaller occupation at large momenta than the equilibrium 
distribution.

\medskip

We display in Table \ref{mome} the momenta $ I_n^{dm} $ defined by 
eq.(\ref{dfIn}) for the normalized DM distributions considered in this paper.

\begin{table}
\begin{tabular}{|c|c|} \hline  
 &  \\
Thermal FD and DW  & $ \chi $ model\\
 &  \\
\hline 
 &  \\
$ I_n^{dm} = \displaystyle \frac2{3 \; \zeta(3)} \; (1 - 2^{-n}) \; n! \; \zeta(n+1) $ & 
$ I_n^{dm} = \displaystyle \frac4{3 \; \zeta(5) \; \sqrt{\pi}} \; 
\Gamma\left(n+\frac12\right) \; \; \zeta(n+3) $ \\
 & \\ \hline  
\end{tabular}
\caption{The normalized momenta $ I_n^{dm} $ defined by eq.(\ref{dfIn}).
Notice that $ I_2^{dm} \equiv 1 $.}
\label{mome}
\end{table}

\medskip

\begin{figure}[h]
\begin{center}
\begin{turn}{-90}
\psfrag{"pin.dat"}{$ \Pi(x) $ vs. $ x $ for a Fermi-Dirac distribution}
\psfrag{"pine.dat"}{$ \Pi(x) $ vs. $ x $ for  sterile neutrinos in the $ \chi $ model}
\includegraphics[height=13.cm,width=8.cm]{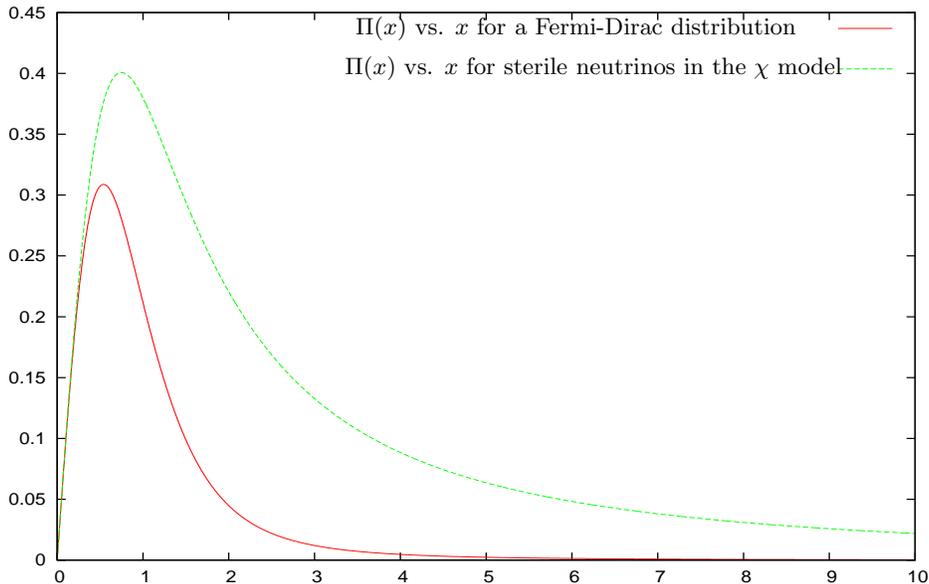}
\end{turn}
\caption{The kernel $ \Pi(x) $ defined by eq.(\ref{defpi}) as a function of $ x $
for fermions in thermal equilibrium (which is identical to the DW model)
and for sterile neutrinos out of thermal equilibrium in the $ \chi $ model.} 
\label{pi}
\end{center}
\end{figure}

For fermions decoupling ultrarelativistically at thermal equilibrium 
(and in the DW model of sterile neutrinos) the normalized freezed out 
distribution function is given 
by eq.(\ref{fdsd}), and the kernel $ \Pi(x) $ for the non-relativistic regime
eq.(\ref{defpi}) can be expressed as
$$
 \Pi^{FD}(x)=\frac{4 \; x}{3 \; \zeta(3)}
  \sum_{n=1}^\infty \frac{(-1)^{n+1} \; n}{(n^2+x^2)^2} \; .
$$
This kernel decreases for large argument $ x $ as
$$
\Pi^{FD}(x)\buildrel{ x \to \infty}\over= \frac1{3\; \zeta(3) \;  x^3}
 +\mathcal{O}\left(\frac1{x^5}\right) \; .
$$
For sterile neutrinos out of thermal equilibrium in the $ \chi $ model the
kernel $ \Pi(x) $ for the non-relativistic regime
eq.(\ref{defpi}) can be expressed as
\be 
\Pi^{\chi}(x)=\frac{\sqrt2}{3 \; \zeta(5)}\sum_{n=1}^\infty\frac{\sqrt{(n^2+x^2)^\frac32 + 3 \; n \; x^2
- n^3 }}{n^\frac52 \; (n^2+x^2)^\frac32 } \quad .
\ee
This kernel decreases for large argument $ x $ as
$$
\Pi^{\chi}(x)\buildrel{ x \to \infty}\over= \frac{\sqrt2 \; \zeta(5/2)}{3 \; \zeta(5)} \; \frac1{x^{3/2}}
 +\mathcal{O}\left(\frac1{x^{5/2}}\right) \; .
$$
We plot $ \Pi^{FD}(x) $ and  $ \Pi^{\chi}(x) $ as functions of $ x $ in fig. \ref{pi}.

\medskip

$ \Pi^{\chi}(x) $ has a longer tail than $ \Pi^{FD}(x) $ due to the higher occupancy of the
low $ Q $ modes in the out of equilibrium momentum distribution. 
The out of equilibrium kernel $ \Pi^{\chi}(x) $ therefore exhibits a longer memory than  $ \Pi^{FD}(x) $.

\section{Volterra integral equations for cold dark matter}\label{apcdm}

All the framework of this paper easily generalizes to cold dark matter (CDM),
that is DM particles with mass beyond one GeV that decouples nonrelativistically.
For CDM the parameter $ \xi_{dm} $ is much larger than for WDM. Typically,
$$
\xi_{cdm} = \frac{m \; a_{eq}}{T_d} =  10^{11} \; \frac{m}{100 \; {\rm GeV}} 
\; \frac{5 \; {\rm GeV}}{T_{d,phys}}
$$
where we take 5 GeV as reference value for the physical decoupling temperature $ T_{d,phys} $ of CDM.
Other values for $ T_{d,phys} $ appears in the literature according to the 
particle physics model chosen but $ \xi_{cdm} $ turns out to be very large 
in all cases (and much larger than for WDM where $ \xi_{wdm} \sim 5000 $).

\medskip

CDM decouples being (by definition) nonrelativistic thus we have at decoupling
$ \varepsilon(y_d,Q) = \xi_{cdm} \; y_d \gg 1 $. In addition, we have 
for the ratio of cdm to radiation densities $ {\cal R}_0(y) = y $
at all times after decoupling.

\medskip

CDM decouples at thermal equilibrium with a normalized Boltzmann distribution
 function \cite{kt,dvs}
\be
f_0^{cdm}(Q) = \sqrt{\frac2{\pi}} \; \frac{e^{-Q^2/(2 \, x)}}{x^\frac32} \quad {\rm where} \quad
x \equiv \xi_{cdm} \; y_d \gg 1 \; .
\ee
For DM particles decoupling nonrelativistic we have instead of eq.(\ref{defDb})
for DM decoupling ultrarelativistic,
\be
{\bar \Delta}_{dm}(y,\kappa)= \xi_{cdm} \; y \; \int \frac{d^3Q}{4 \, \pi} \;
f_0^{dm}(Q) \; \frac{\Psi_{dm}(y, \vQ, \vka)}{\psi(y_d,\vka)} \quad ,
\ee
and therefore at the initial (decoupling) time $ y = y_d $
\be\label{dcdm0}
{\bar \Delta}_{dm}(y_d,\kappa)= x \; \int_0^{\infty}  Q^2 \; dQ
\; f_0^{dm}(Q) \; {\bar c}_{dm}^0(Q) \; ,
\ee
where we used the initial conditions for the BV distribution function
discussed in ref.\cite{uno}
$$
\Psi_{dm}(y_d,\vQ,\vka) = \psi(y_d,\vka) \; {\bar c}_{dm}^0(q) \; .
$$
From eqs.(\ref{condini}) and (\ref{dcdm0}) we find as initial CDM fluctuations for TIC 
$$
{\bar \Delta}_{cdm}(y_d,\kappa) = \frac12 \; \xi_{cdm} \; y_d \; 
\int_0^{\infty}Q^3 \; dQ \; \frac{df_0^{dm}}{dQ} = -\frac32 \; x \; .
$$
For CDM the free-streaming distance is defined similarly to 
eq.(\ref{lfsI}) as
\be\label{lcdm}
\int_{y_d}^{y} \frac{dy'}{\sqrt{(1+y') \; \left[ y'^2 + \displaystyle 
\left(\displaystyle Q/\xi_{cdm}\right)^2 \right]}} \simeq \int_{y_d}^{y} 
\frac{dy'}{y' \; \sqrt{1+y'}} = l_{cdm}(y) \quad,\quad {\rm thus,} \quad 
l_{cdm}(y) = \ln \left(\frac4{y_d}\right) + 2 \, s(y) \; .
\ee
The free-streaming distance turns to be independent of $ Q $ as it should be because 
CDM particles are very slow.

\medskip

Let us consider zero anisotropic stress here, for simplicity. Thus, the evolution
of the CDM fluctuations is given by the single Volterra equation (\ref{sinneu}).
Since CDM particles are always nonrelativistic with $ \xi_{cdm} \; y \gg 1 $ the coefficients
and kernel in eq.(\ref{sinneu}) take the form
\bea\label{abcdm}
&& a_{cdm}(y,\alpha)= 
\frac{2 \, \xi_{cdm} \; y}{\alpha \; l(y)} \int_0^{\infty}  Q \; dQ
\left[ f_0^{dm}(Q) \; {\bar c}_{dm}^0(Q)+\frac{df_0^{dm}}{d\ln Q}\right]
\; \sin \left[\frac{\alpha}2 \, Q \, l(y) \right] \; , \cr \cr\cr
&& b_{cdm}(y) = 3 \quad , \quad
N_\alpha(y,y') = - (\xi_{cdm})^2 \; y \; y' \; \Pi\left[ \alpha \; 
\left( s(y)-s(y')\right)\right] \; ,
\eea
where we used eqs.(\ref{asd3}), (\ref{dfb}), and (\ref{Nnr}).
The Volterra integral equation for CDM takes thus a form similar to 
eq.(\ref{sinneu}) for WDM
\bea\label{Vcdm}
&& {\bar \Delta}_{cdm}(y,\alpha)= a_{cdm}(y,\alpha) + 3 \; \xi_{dm} \; y \; {\breve \phi}(y,\alpha)
+ \kappa \; \int_{y_d}^y \frac{dy'}{\sqrt{1+y'}} \; N_\alpha(y,y')\; 
{\breve \phi}(y',\alpha)  \quad {\rm and ~ also} \\ \cr
&&  {\bar \Delta}_{cdm}(y,\alpha)= a_{cdm}(y,\alpha) + 3 \; \xi_{dm} \; y \; {\breve \phi}(y,\alpha)
- 2 \,  (\xi_{cdm})^2 \; \kappa \; y \; \int_{s(y_d)}^{s(y)} \frac{ds'}{\sinh^4 s'} \; 
\Pi\left[\alpha\left(s(y)-s'\right)\right] \;{\breve \phi}(y'=\frac1{\sinh^2 s'},\alpha)  \quad . \nonumber
\eea
At $ y = y_d , \; l_{cdm}(y_d) = 0 $ and eqs.(\ref{dcdm0}) and (\ref{abcdm}) yield
$$
a_{cdm}(y_d,\alpha)= {\bar \Delta}_{cdm}(y_d,\kappa) - 3 \; x \; .
$$
Therefore, the Volterra equation (\ref{Vcdm})
is identically satisfied at $ y = y_d $ since $ {\breve \phi}(y_d,\alpha) = 1 $.

\medskip

The whole section \ref{solk0} translates to the CDM case. Since
for CDM $ \xi_{cdm} \; y \gg 1 $ eq.(\ref{fnr}) results for all $ y $
$$
{\bar \delta}_{cdm} (y, 0) = 3 \; {\bar \Theta}_{r,0} (y, 0) \; ,
$$
as it must be.

\acknowledgments

We are grateful to D. Boyanovsky and C. Destri for useful discussions.

\appendix

\section{The gravitational potential in the RD era}\label{potr}

During the RD era the gravitational potential $ \phi(\eta,\va) $ is dominated by the 
radiation (photons and neutrino) fluctuations. Neglecting the anisotropic stress, the following equations 
relate the gravitational potential with the first two radiation momenta \cite{dod,mab}
\bea\label{3ecs}
&& -k^2 \; \phi(\eta,\va) = 16 \, \pi \; G \; a^2(\eta) \; \rho_r(\eta)
\left[ \Theta_{r,0}(\eta,\va) + \frac3{k} \; h(\eta) \;  \Theta_{r,1}(\eta,\va)
\right] \; , \cr \cr 
&& \frac{d\Theta_{r,0}}{d \eta}  + k \; \Theta_{r,1}(\eta,\va) 
= \frac{d\phi}{d \eta} \; , \cr \cr 
&& \frac{d\Theta_{r,1}}{d \eta} - \frac{k}3 \; \Theta_{r,0}(\eta,\va)= 
\frac{k}3 \;\phi(\eta,\va) \; .
\eea
Here $ \Theta_{r,0}(\eta,\va) $ and $ \Theta_{r,1}(\eta,\va) $ are the first two momenta
of the radiation temperature field
$$
\Theta_{r,0}(\eta,\va) = R_\gamma(\eta) \; \Theta_0(\eta,\va) +  R_\nu(\eta) \; N_0(\eta,\va)
\quad {\rm and} \quad  N_0(\eta,\va) = \frac1{4 \; I_3^\nu} \; {\bar \Delta}_{\nu}(y,\kappa) \; \phi(0,\va) \; .
$$
The infinite hierarchy of equations 
arising from the Boltzmann-Vlasov equation for radiation and matter,
has been truncated to the first two equations \cite{dod,mab}. 
$ h(\eta) = d \ln a/d \eta $ stands for the Hubble parameter and 
$ \rho_r(\eta) =  \Omega_r \; \rho_c / a^4(\eta) $ for the radiation density.

\medskip

Eliminating $ \Theta_{r,0}(\eta,\va) $ among eqs.(\ref{3ecs}) yields
\bea\label{2eta}
&& \frac{d}{d \eta} \left[ h(\eta) \; \Theta_{r,1}(\eta,\va) \right]- \frac{k^2}3 \;
\Theta_{r,1}(\eta,\va) -  \frac{k}3 \; \frac{d\phi}{d \eta}(\eta,\va) 
- \frac{k^2}{48 \; \pi \; G} \; \frac{d}{d \eta}
\left[\frac{\phi(\eta,\va)}{a^2(\eta) \; \rho_r(\eta)}\right] = 0 \; , \cr \cr 
&&  \frac{d\Theta_{r,1}}{d \eta} +  h(\eta) \; \Theta_{r,1}(\eta,\va)
+ \frac{k}3 \left[1 - \frac{k^2}{16 \, \pi \; G \; a^2(\eta) \; \rho_r(\eta)}\right]
\phi(\eta,\va) = 0 \; .
\eea
It is convenient to use the variable $ y $ defined in eq.(\ref{ay}) instead of the conformal
time $ \eta $. We find in terms of $ y $
\be\label{fotones}
a^2(\eta) \; \rho_r(\eta) = \frac3{8 \, \pi \; G} \; \frac1{{\eta^*}^2 \; y^2}
\ee
and eqs.(\ref{2eta}) read
\bea\label{2ecs}
&&\frac{d\Theta_{r,1}}{d y}- \frac1{1+y} \left(\frac1{y} + \frac12 + \frac{\kappa^2}3
\; y \right)\Theta_{r,1}(y,\va) = \frac{\kappa}{3 \; \sqrt{1+y}} \left[ \left(1 + 
\frac{\kappa^2 \; y^2}6  \right)y \; \frac{d}{d y} + \frac{\kappa^2 \; y^2}3
\right]\phi(y,\va)  \; , \cr \cr 
&& 
\frac{d\Theta_{r,1}}{d y} + \frac1{y} \; \Theta_{r,1}(y,\va)= -\frac{\kappa}{3 \; \sqrt{1+y}}
\left(1 - \frac{\kappa^2 \; y^2}6  \right) \; \phi(y,\va) \; .
\eea
Eliminating now $ \Theta_{r,1}(\eta,\va) $ we have
\be\label{otraec}
\frac2{y} \left(1 + \frac34 \, y + \frac{\kappa^2 \; y^2}6 \right)\Theta_{r,1}(y,\va)+
\frac{\kappa}3 \; \sqrt{1+y}  \left(1 +\frac{\kappa^2 \; y^2}6 \right)\left[1+
y \; \frac{d}{d y}\right]\phi(y,\va)=0 \quad .
\ee
Taking the $ y $ derivative of this equation and replacing $ d\Theta_{r,1}(y,\va)/dy $ 
and $ \Theta_{r,1}(y,\va) $ from eqs.(\ref{2ecs}) and (\ref{otraec}) respectively, we get the
second order differential equation for the gravitational potential $ \phi(\eta,\va) $:
\be\label{1ec}
\frac{d^2\phi}{d y^2} + \frac2{y} \; R_\kappa(y) \; \frac{d\phi}{d y}
 + \frac2{y^2} \;  S_\kappa(y) \; \phi(y,\va)=0 \; ,
\ee
where,
\bea
&& R_\kappa(y) \equiv 1+ \frac{y}{4 \; (1+y)}
+ \frac{\displaystyle \kappa^2 \; y^2/6}{1 + \displaystyle \kappa^2 \; y^2/6}+ 
 \frac{1 + \frac38 \; y}{1 + \frac34 \; y+ \displaystyle \kappa^2 \; y^2/6}  \cr \cr\cr
%&& R_\kappa(y) \equiv 1+ \frac{y}{4 \; (1+y)}
%+ \frac{\displaystyle\frac{\kappa^2 \; y^2}6}{1 + \displaystyle\frac{\kappa^2 \; y^2}6}+ 
% \frac{1 + \frac38 \; y}{1 + \frac34 \; y+ \displaystyle\frac{\kappa^2 \; y^2}6}  \cr \cr\cr
%&& S_\kappa(y) \equiv\frac{\displaystyle\frac{\kappa^2 \; y^2}6}{\displaystyle 1 
%+ \frac{\kappa^2 \; y^2}6} +
%\displaystyle\frac{1 + \displaystyle\frac38 \; y}{\displaystyle 1 
%+ \frac34 \; y+ \frac{\kappa^2 \; y^2}6}
%-\frac{\displaystyle 1 +\frac{y}2 \; \left(1 - \displaystyle \frac{\kappa^2 \; y^2}6\right)-
%\displaystyle \frac{\kappa^4 \; y^4}{36}}{(1+y) \; \left(1 + 
%\displaystyle\frac{\kappa^2 \; y^2}6\right)} \; .
&& S_\kappa(y) \equiv\frac{\displaystyle \kappa^2 \; y^2/6}{\displaystyle 1 
+ \kappa^2 \; y^2/6} +
\displaystyle\frac{1 + \displaystyle\frac38 \; y}{\displaystyle 1 
+ \frac34 \; y+ \kappa^2 \; y^2/6}
-\frac{\displaystyle 1 +\frac{y}2 \; \left(1 - \displaystyle \kappa^2 \; y^2/6\right)-
\displaystyle \kappa^4 \; y^4/36}{(1+y) \; \left(1 + 
\displaystyle \kappa^2 \; y^2/6\right)} \; .
\eea

\begin{figure}[h]
\begin{center}
\begin{turn}{-90}
\psfrag{"firzn01.dat"}{${\breve\phi}(y,\alpha=0.1)$}
\psfrag{"firzn1.dat"}{${\breve\phi}(y,\alpha=1)$}
\psfrag{"firzn10.dat"}{${\breve\phi}(y,\alpha=10)$}
\psfrag{"j1zn1.dat"}{$\phi^0(\zeta)$}
\includegraphics[height=17.cm,width=10.cm]{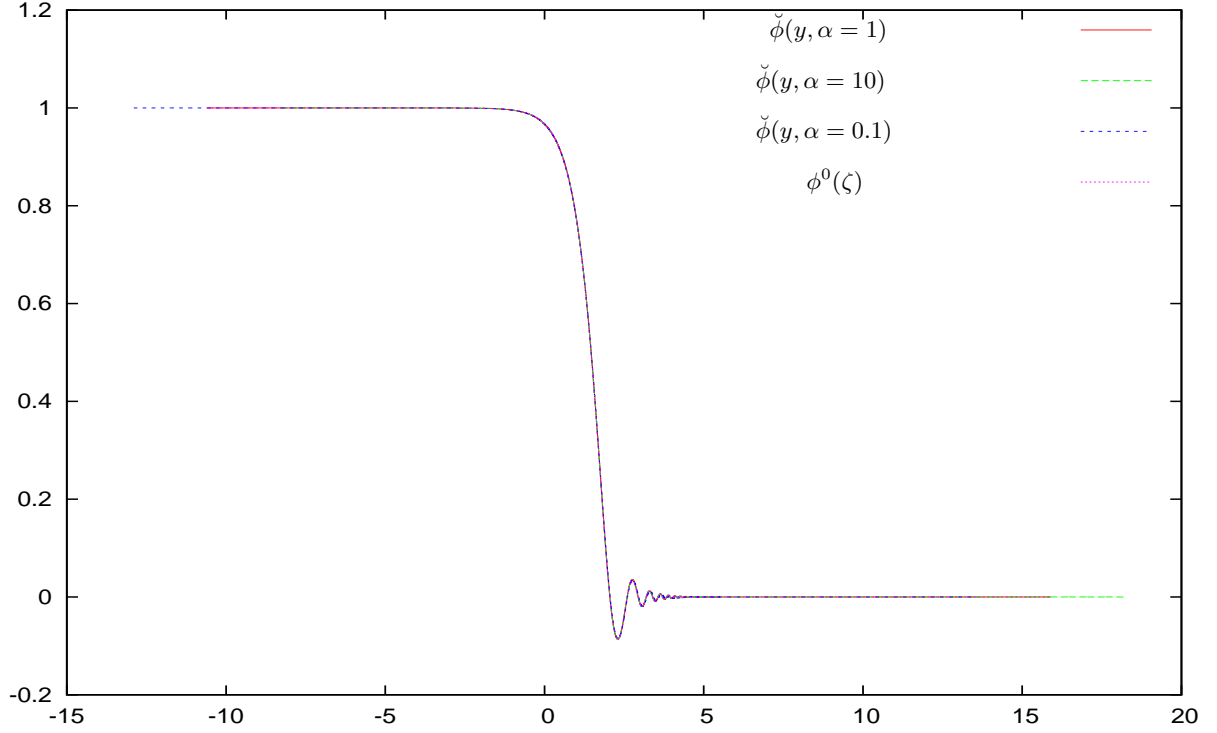}
\end{turn}
\caption{The gravitational potential $ {\breve\phi}(y,\alpha) $ 
 and $ \phi^0(y,\alpha) $ vs.  $ \log_{10} \zeta $ for $ \alpha = 0.1, 1 $ and 10 
defined by eqs. (\ref{firsom}) and (\ref{firsom0}). We see that $ \phi^0(\zeta) $
is a very good approximation to $ {\breve\phi}(y,\alpha) $ in the whole
range of $ \zeta $.} 
\label{fir}
\end{center}
\end{figure}

In the radiation dominated era and for $ \kappa y \gg 1 $ this equation reduces to 
$$
\frac{d^2\phi}{d y^2} +\frac4{y} \frac{d\phi}{d y}+ \frac13 \; \kappa^2 \; 
\phi(y,\va)=0
$$
whose solution in terms of Bessel functions is given by eq.(\ref{firad}).

\medskip

We solve numerically eq.(\ref{1ec}) both in the radiation dominated
and in the matter dominated eras. We plot in fig. \ref{fir} the normalized gravitational potential
\be\label{firsom}
{\breve\phi}(y,\alpha) \equiv \frac{\phi(y,\va)}{\phi_{prim}(\va)}
\quad , \quad {\breve\phi}(0,\alpha) = 1 \; ,
\ee
as a function of $ \log_{10} \zeta $ for $ \alpha = 0.1, 1 $ and 10 as well as 
the function 
\be\label{firsom0}
\phi^0(\zeta) \equiv 3 \;  \frac{j_1\left(\zeta\right)}{\zeta} 
\quad , \quad \zeta = \frac{\kappa \; y}{\sqrt3} \quad , \quad\phi^0(0)=1
\; .
%\phi(y,\va) = 3 \; \phi_{prim}(\va) \; \frac{\sqrt3}{\kappa \; y} \; 
%j_1\left(\frac{\kappa \; y}{\sqrt3}\right) \; .
\ee
We see from fig. \ref{fir} that the function $ \phi^0(\zeta) $
is a very good approximation to $ {\breve\phi}(y,\alpha) $.

\section{The free-streaming length in the different regimes.}\label{apD}

We evaluate now the integral eq.(\ref{lfsI})
\be\label{rosdA}
l(y,Q) =  \int_0^{y} \frac{dy'}{\sqrt{(1+y') \; 
\left[ y'^2 + \displaystyle \left(\displaystyle Q/\xi_{dm} \right)^2 \right]}} \; ,
\ee
in the different regimes depicted in Table III. This is an elliptic integral 
that can be expressed in terms of the standard incomplete elliptic integrals
of first kind \cite{smi}
\be\label{lfsexa}
l(y,Q) = \left[1 + \left(\displaystyle\frac{Q}{\xi_{dm}}\right)^2\right]^{-\frac14}
\; \left[F(\varphi(0),{\hat p}) - F(\varphi(y),{\hat p})\right] \; ,
\ee
where
$$
2 \; {\hat p}^2 = 1 + \frac1{\sqrt{1 + \left(\displaystyle Q/\xi_{dm} \right)^2}}
\quad , \quad \frac2{\sin \varphi(y)} = \frac{\left[\displaystyle 1 + \left(\displaystyle 
Q / \xi_{dm} \right)^2\right]^\frac14}{\displaystyle \sqrt{1+y}}
+ \displaystyle\frac{\sqrt{1+y}}{\left[1 + \displaystyle\left( Q/\xi_{dm} \right)^2\right]^\frac14}
$$
%but
Taking into account that $ \xi_{dm} \sim 5000 \gg 1 $ we can express this integral quite acurately
in terms of elementary functions.
\begin{itemize}
\item{For $ y < 0.01 \ll 1 $ we can expand $ 1/\sqrt{1+y'} $ in 
eq.(\ref{rosdA}) in powers of $ y' $ and obtain
\bea\label{aprox1}
&&l(y,Q) = \xi_{dm} \; \int_0^{y} \frac{dy'}{\sqrt{Q^2 + (\xi_{dm})^2 \; y'^2}} 
\left[1 - \frac12 \; y' + \frac38 \; y'^2 + {\cal O}\left(y'^3\right) \right]= \cr \cr 
&&= \left[1 -\frac3{16} \; \left(\frac{Q}{\xi_{dm}}\right)^2 \right]
\displaystyle {\rm Arg \, Sinh}\left(\displaystyle\frac{\xi_{dm} \; y}{Q}\right) 
- \frac12 \; \left[\left(1-\frac38 \; y \right)\sqrt{y^2 + \left(\frac{Q}{\xi_{dm}}\right)^2}
- \frac{Q}{\xi_{dm}}\right] + {\cal O}(y^3) \; .
\eea
Notice that in this range of $ y $ and for typical $ Q \sim 1 $ the arguments in 
eq.(\ref{aprox1}) can go from $ Q \gg \xi_{dm} \; y $ till $ Q \ll \xi_{dm} \; y $.
In the case $ \xi_{dm} \; y \ll Q \ll \xi_{dm} $ this formula simplifies as
$$
l(y,Q) \simeq \frac{\xi_{dm} \; y}{Q}  \; .
$$}
\item{For $ y > 0.01 $ it is convenient to split the integral  eq.(\ref{rosdA}) in two pieces:
\be\label{lygran}
l(y,Q) = l(\infty,Q) - \xi_{dm} \; \int_{y}^{\infty} 
\frac{dy'}{\sqrt{(1+y')[Q^2 + (\xi_{dm})^2 \; y'^2]}} \; ,
\ee
where 
$$
l(\infty,Q) = \xi_{dm} \; \int_0^{\infty} 
\frac{dy}{\sqrt{(1+y)[Q^2 + (\xi_{dm})^2 \; y^2]}} \; .
$$
In order to obtain the asymptotic expansion of $ l(\infty,Q) $ 
for $ Q/\xi_{dm} \to 0 $ it is convenient to change the integration
variable as
$$
y = t - 1 - \left(\frac{Q}{2 \, \xi_{dm}}\right)^2 \; \frac1{t-1} \; ,
$$
and $ l(\infty,Q) $ becomes
\be\label{linfi}
l(\infty,Q) =\int_{1 + \frac{Q}{2 \, \xi_{dm}}}^{\infty} 
\frac{dt}{\sqrt{t} \; (t-1)} \; \frac1{\sqrt{1 -
\displaystyle\frac{\left(Q/\xi_{dm}\right)^2}{4 \; t \; (t-1)}}} \quad .
\ee
Expanding the integrand of eq.(\ref{linfi}) in powers of $ (Q/\xi_{dm}) ^2 $
and integrating term by term yields the asymptotic expansion
\be\label{linfi2}
l(\infty,Q) = \left[1-\frac3{16} \; \left(\frac{Q}{\xi_{dm}}\right)^2
\right] \log\left(\displaystyle \frac{8 \; \xi_{dm}}{Q}\right)
+ \frac12 \; \frac{Q}{\xi_{dm}} + \frac7{16} \; \left(\frac{Q}{\xi_{dm}}\right)^2 + 
{\cal O}\left(\left[\frac{Q}{\xi_{dm}}\right]^3\right) \; .
\ee
The integral in the second term of eq.(\ref{lygran}) is evaluated
by expanding the integrand in powers of $ (Q/\xi_{dm})^2 $
\bea\label{linfi3}
&& \xi_{dm} \; \int_{y}^{\infty} 
\frac{dy'}{\sqrt{(1+y')[Q^2 + (\xi_{dm})^2 \; y'^2]}} =
\int_{y}^{\infty}\frac{dy'}{y' \; \sqrt{1+y'}}-\frac12 \;
\left(\frac{Q}{\xi_{dm}}\right)^2 \int_{y}^{\infty}\frac{dy'}{y'^3
 \; \sqrt{1+y'}} + {\cal O}\left(\left[\frac{Q}{\xi_{dm}}\right]^3\right)=
\cr \cr \cr
&& = 2 \, \left[1 -\frac3{16} \; \left(\frac{Q}{\xi_{dm}}\right)^2 \right]
\; {\rm Arg \, Sinh}\left(\frac1{\sqrt{y}} \right) 
+ \frac18 \; \left(\frac{Q}{y \; \xi_{dm}}\right)^2
\left[3 \, y \; \sqrt{1+y} + y
+ 2 \right]+{\cal O}\left(\left[\frac{Q}{\xi_{dm}}\right]^3\right)\; .
\eea
Collecting together eqs.(\ref{lygran}), (\ref{linfi2}) and (\ref{linfi3}) yields
\bea \label{roapr}
&& l(y,Q) = \left[1 -\frac3{16} \; \left(\frac{Q}{\xi_{dm}}\right)^2 \right]
\left[\log\left(\displaystyle \frac{8 \; \xi_{dm}}{Q}\right)-
2 \, {\rm Arg \, Sinh}\left(\frac1{\sqrt{y}} \right) \right]+\cr \cr
&& + \frac12 \; \frac{Q}{\xi_{dm}} + \frac7{16} \; 
\left(\frac{Q}{\xi_{dm}}\right)^2 - \frac18 \; \left(\frac{Q}{y \; \xi_{dm}}\right)^2
\left[3 \, y \; \sqrt{1+y} + y
+ 2 \right]+{\cal O}\left(\left[\frac{Q}{\xi_{dm}}\right]^3\right)\; .
\eea
}
\item{For $ y \gg 1 $, eq.(\ref{roapr}) becomes
$$
l(y,Q) \simeq \displaystyle 
\log\left(\displaystyle\frac{8 \; \xi_{dm}}{Q}\right)+ \frac12 \; \frac{Q}{\xi_{dm}}
- \displaystyle\frac2{\sqrt{y}}  +{\cal O}\left(\left[\frac{Q}{\xi_{dm}}\right]^2
\log\left(\displaystyle\frac{8 \; \xi_{dm}}{Q}\right)\right)
\; .$$}
\end{itemize}
We have thus derived the four entries for $ l(y,Q) $ in Table III.

\medskip

Moreover, ref. \cite{ks} provides asymptotic expressions for the incomplete elliptic
integral of first kind $ F(\varphi(y),k) $ valid for $ k'^2 \ll 1 $ (therefore $ Q \ll \xi_{dm} $)
which are uniform in $ \varphi $. The first order asymptotic expression takes the
form \cite{ks}
\bea\label{eliplio}
&& F_1(\varphi(y),k) = \left[1+\frac18\left(1-\frac1{p}\right)\right] \; 
\log\left|\frac{\sqrt{u}+\sqrt{p}}{\sqrt{u}-\sqrt{p}}\right| - \frac12 \;
\left(\sqrt{\frac{u}{p}} + \sqrt{\frac{p}{u}} \right) \; 
\log\left[\frac12+\frac{\sqrt{\displaystyle y^2 +  
\left(Q/\xi_{dm}\right)^2}}{2 \; \displaystyle|u-p|}\right] \; , \cr \cr
&& p \equiv \sqrt{1 + \left(Q/\xi_{dm}\right)^2} \quad , \quad u \equiv 1 + y \; . 
\eea
We display in fig. \ref{roL}  $ l(y,Q) $ 
%from  the exact eq.(\ref{rosdA}) and
%from the approximation formulas eq.(\ref{aprox1}) and (\ref{roapr}) 
for $ Q = 0.1, \; 1 $ and $ 10 $.
%We see  that both approximations are excellent in their range of validity.
%Similar results are obtained for relevant values of $ Q \lesssim 50 $.
Large values of $ Q $ get suppressed in the integrals by the decrease of
the distribution function $ f_0^{dm}(Q) $.


\begin{thebibliography}{}
\bibitem{uno} H. J. de Vega, N. G. Sanchez, accompanying paper 
`Cosmological evolution of warm dark matter fluctuations I: 
Efficient computational framework with Volterra integral equations',
ArXiv:1111.0290.
\bibitem{dod} Dodelson S, \textit{Modern Cosmology},  
Academic Press, 2003.
\bibitem{mab} C-P. Ma, E. Bertschinger, ApJ, 455, 7 (1995).
\bibitem{sz} 
U. Seljak,  ApJ 435, L87 (1994), U. Seljak, M. Zaldarriaga,  ApJ  469, 437 (1996).
\bibitem{dw} S. Dodelson, L. M. Widrow, Phys. Rev. Lett. \textbf{72}, 17 (1994).
\bibitem{neus} X. Shi, G. M. Fuller, Phys. Rev. Lett. \textbf{82}, 2832 (1999).
\bibitem{neus2} 
A. D. Dolgov,	Phys. Rept. 370 (2002) 333-535.
F. Munyaneza, P. L. Biermann, Astron. and Astrophys., 458, L9 (2006).
A. Kusenko, Phys. Rept. 481, 1 (2009).
\bibitem{mas1} K. Abazajian, G. M. Fuller, M. Patel, Phys. Rev.
\textbf{D64}, 023501 (2001);  K. Abazajian, G. M. Fuller, Phys. Rev.
\textbf{D66}, 023526, (2002); G. M. Fuller \emph{et. al.}, Phys. Rev.
\textbf{D68}, 103002 (2003); K. Abazajian, Phys. Rev. \textbf{D73}, 063506  (2006).
P. L. Biermann, A. Kusenko, Phys. Rev. Lett. \textbf{96}, 091301 (2006);
T. Asaka, M. Shaposhnikov, A. Kusenko; Phys. Lett. \textbf{B638}, 401 (2006).
\bibitem{dvs} H. J. de Vega, N. G. S\'anchez, arXiv:0901.0922, 
Mon. Not. R. Astron. Soc. {\bf 404}, 885 (2010).
\bibitem{kt}  Kolb EW and Turner MS, \textit{The Early
Universe}, Addison Wesley. Redwood City, C.A. 1990.
\bibitem{gil} I. H. Gilbert, Astrophys. J. \textbf{144}, 233
(1966); \emph{ibid}, \textbf{152}, 1043 (1968).
\bibitem{BBKS} J M Bardeen et al. Astrophys. J. \textbf{304}, 15 (1986).
\bibitem{turokWDM} P. Bode, J. P. Ostriker, N. Turok, Astrophys. J \textbf{556}, 93 (2001). 
K. Abazajian, Phys. Rev. {\bf D 73}, 063513 (2006).
\bibitem{bdvs} D. Boyanovsky, H. J. de Vega, N. G. Sanchez,
Phys. Rev. {\bf D 78}, 063546 (2008).
\bibitem{biblia} D. Boyanovsky,  C. Destri,  H. J. de Vega, N. G. S\'anchez,  
arXiv:0901.0549, Int. J. Mod. Phys. {\bf A 24}, 3669-3864 (2009).
\bibitem{eih} A. Einstein, L. Infeld, B. Hoffmann, Annals of Mathematics,
{\bf 39}, 65 (1938).
\bibitem{WMAP}
E. Komatsu et al. (WMAP collaboration), Astrophys. J. Suppl. 180:330 (2009).
\bibitem{bwu} D. Boyanovsky, J. Wu, Phys. Rev. D83, 043524 (2011). 
\bibitem{smi} V. I. Smirnov, Course of Higher Mathematics vol. III, part 2, Pergamon Press,
1964, Oxford. 
\bibitem{ks} D. Karp, S. M. Sitnik, J. of Computational and Appl. Math. 205, 186 (2007).
\bibitem{modelos} M. Shaposhnikov, I. Tkachev, Phys. Lett. B639, 414 (2006).
\bibitem{fuera} A. Kusenko, PRL 97, 241301 (2006).
K. Petraki,  A. Kusenko, Phys. Rev. D77, 065014 (2008).
K. Petraki,  Phys. Rev. D77, 105004 (2008).
D. Boyanovsky, J. Wu, Phys. Rev. D83, 043524 (2011). 
\bibitem{distest} D. Boyanovsky, Phys. Rev. D78:103505, (2008).
\bibitem{als} T. Asaka, M. Laine, M. Shaposhnikov, 
	JHEP 0701:091,2007.
\bibitem{ddv}  D. Boyanovsky, C. Destri, H. J. de Vega, Phys. Rev. \textbf{D69}, 045003 (2004).

C. Destri, H. J. de Vega, Phys. Rev. \textbf{D73}, 025014 (2006).
\end{thebibliography}
\end{document}